\documentclass[aps,prb,twocolumn,preprintnumbers,amsmath,amssymb,superscriptaddress]{revtex4-2}

\usepackage{tabularx}
\usepackage[utf8]{inputenc} 
\usepackage[T1]{fontenc}    
\usepackage{hyperref}       
\usepackage{url}            
\usepackage{amsfonts}       
\usepackage{amsmath,amsthm,amssymb} 
\usepackage{bm}
\usepackage{bbm}
\usepackage{dcolumn}
\usepackage{nicefrac}       
\usepackage{gensymb}
\usepackage{graphicx}
\usepackage{xcolor, etoolbox}
\usepackage{mathtools}
\usepackage{mathrsfs}
\usepackage{ragged2e}
\usepackage{microtype}      
\usepackage{braket}      
\usepackage{enumitem}
\usepackage{xspace}
\usepackage{dsfont}
\usepackage{float}
\usepackage{soul}
\usepackage{orcidlink}

\DeclarePairedDelimiter\abs{\lvert}{\rvert}

\newcommand\varpm{\mathbin{\vcenter{\hbox{%
  \oalign{\hfil$\scriptstyle+$\hfil\cr
          \noalign{\kern-.3ex}
          $\scriptscriptstyle({-})$\cr}%
}}}}
\newcommand\varmp{\mathbin{\vcenter{\hbox{%
  \oalign{$\scriptstyle({+})$\cr
          \noalign{\kern-.3ex}
          \hfil$\scriptscriptstyle-$\hfil\cr}%
}}}}

\hypersetup{
    colorlinks,
    linkcolor=blue,
    citecolor=blue,
    urlcolor=blue,
    breaklinks=true 
}

\begin{document}

\title{Relativistic Mott transitions and finite-temperature effects\\ of quantum criticality in Dirac semimetals}

\author{Mireia Tolosa-Sime\'on\,\orcidlink{0000-0003-4032-0132}}
\email{tolosa@tp3.rub.de}
\affiliation{Theoretische Physik III, Ruhr-Universit\"at Bochum, D-44801 Bochum, Germany}

\author{Laura Classen\, \orcidlink{0000-0002-3513-9403}}
\email{l.classen@fkf.mpg.de}
\affiliation{Max Planck Institute for Solid State Research, D-70569 Stuttgart, Germany}
\affiliation{Department of Physics, Technical University of Munich, D-85748 Garching, Germany}

\author{Michael M. Scherer\, \orcidlink{0000-0003-0766-9949}}
\email{scherer@tp3.rub.de}
\affiliation{Theoretische Physik III, Ruhr-Universit\"at Bochum, D-44801 Bochum, Germany}

\begin{abstract}
Gross--Neveu--Yukawa-type models such as the chiral Ising, chiral XY, and chiral Heisenberg models, serve as effective descriptions of two-dimensional Dirac semimetals undergoing quantum phase transitions into various symmetry-broken ordered states. 
Their relativistic quantum critical points govern the systems' physical behavior in the vicinity of the transition  
also at finite temperatures, which is strongly influenced by critical order-parameter and chiral fermion fluctuations.
Here, we explore the effect of these fluctuations at zero and finite temperature, both in the Dirac phase and in the Mott phases with spontaneously broken symmetry.
To that end, we set up a functional renormalization group approach, which allows us to systematically calculate the quantum phase diagrams and scaling behavior at and near quantum criticality. 
We explicitly estimate quantum critical exponents, calculate the quasiparticle weight of the chiral Dirac excitations, and determine the extent of the quantum critical fan. Furthermore, we 
expose a semimetallic precondensation regime where order-parameter fluctuations destroy order at finite temperature and we show the related manifestation of the Coleman--Hohenberg--Mermin--Wagner theorem. 
For the chiral XY model, we also expose signatures of Berezinskii--Kosterlitz--Thouless physics in a system that includes strong fermion fluctuations. 
In view of recent experimental developments on correlated phases in highly tunable two-dimensional Dirac materials, our work aims at a more comprehensive theoretical description of relativistic quantum criticality in semimetals, including non-Dirac liquid behavior. 
\end{abstract}

\maketitle

\section{Introduction}
The advent of Dirac materials~\cite{Wehling_2014,Vafek_2014} has fueled the interest in their quantum phase transitions driven by strong correlations~\cite{Boyack:2020xpe}.
Due to the presence of gapless Dirac excitations, such transitions cannot be described by order parameter fluctuations alone; hence, they challenge the Landau--Ginzburg--Wilson paradigm~\cite{Landau:1980mil}.
Naturally, quantum phase transitions are a zero-temperature phenomenon driven by the variation of a non-thermal control parameter. 
However, the point of a continuous quantum phase transition, i.e., the quantum critical point (QCP), can impact observables over a wide range of the phase diagram, even at finite temperatures~\cite{Vojta_2003,Sachdev2011}. 
Around metallic quantum critical points this does not only include bosonic excitations of the order parameter, but fermion excitations also show anomalous behavior inducing a non-Fermi liquid regime \cite{doi:10.1080/001075199181602,RevModPhys.73.797,RevModPhys.79.1015,doi:10.1063/1.3554314,RevModPhys.88.025006,doi:10.1146/annurev-conmatphys-031016-025531,doi:10.1146/annurev-conmatphys-031218-013339,Xu_2019,RevModPhys.94.035004}. 
In addition, gapless fermion modes fundamentally change the universal critical behavior by providing additional source fluctuations beyond those coming from the order-parameter.
In the particular case of Dirac materials undergoing a continuous quantum phase transition, it has been argued that their quantum critical behavior can be captured by  Gross-Neveu-type models that are well known in the context of relativistic quantum field theory~\cite{Herbut2006,Herbut2009,Boyack:2020xpe,Herbut:2023xgz}.

Experimental efforts notwithstanding, the prototypical Dirac material --  graphene --  could not be observed to exhibit such behavior. 
Yet, the discovery of correlated phases in twisted bilayer graphene near the magic angle~\cite{cao2018correlated,cao2018unconventional} has inspired theoretical proposals, suggesting that quantum criticality of Dirac electrons can be accessed by twist-angle tuning in such Van-der-Waals heterostructures~\cite{Parthenios2023,huang2024angletunedgrossneveuquantumcriticality,biedermann2024twisttunedquantumcriticalitymoire}.
Similarly, the recent experimental observation of a  transition from a Dirac semimetal into an insulating state in twisted WSe${}_2$ tetralayers by variation of the twist angle~\cite{ma2024relativisticmotttransitionstrongly} can be expected to fall into the universality class of Dirac quantum criticality.
 Also, certain organic crystals host massless Dirac fermions which can undergo an interaction-induced transition to insulating behavior~\cite{PhysRevLett.116.226401,Hirata_2016,doi:10.1126/science.aan5351}.

Typical examples for insulating ordered phases that have been argued to be present in strongly-correlated Dirac materials with Fermi level at the Dirac point include charge-density wave (CDW) or valley Hall order~\cite{Herbut2006,Raghu2008,Weeks2010,Grushin2013,PhysRevX.10.031034,Classen_2022,PhysRevX.12.011061}, Kekul\'e valence-bond-solid order~\cite{PhysRevB.62.2806,PhysRevLett.98.186809}, intervalley-coherent (IVC) states~\cite{PhysRevX.8.031089,PhysRevX.10.031034,PhysRevX.11.011014}, or antiferromagnetism~\cite{sorella1992semi,Herbut2006,PhysRevLett.100.146404,PhysRevResearch.5.043173} (AFM). 
Employing zero-temperature field theory, there is extensive work describing the quantum critical behavior of these transitions in terms of effective Gross--Neveu--Yukawa-type models, e.g., the chiral Ising, chiral XY, and chiral Heisenberg models~\cite{Herbut2006,Boyack:2020xpe,Parthenios2023,Herbut:2023xgz}. 
The field-theoretical estimates for the critical exponents agree well with the conformal bootstrap~\cite{Iliesiu:2015qra,Iliesiu:2017nrv,Erramilli2023} and quantum Monte Carlo~(QMC) simulations~\cite{PhysRevD.88.021701,PhysRevB.101.064308,PhysRevB.108.L121112} for the chiral Ising model.
For the chiral XY model the agreement with QMC is also very reasonable~\cite{LiFermi:2017,Otsuka:2018,huang2024angletunedgrossneveuquantumcriticality,Hawashin2025}, while the quantum-critical exponents for the chiral Heisenberg model are still under intense investigation~\cite{PhysRevX.3.031010,PhysRevB.91.165108,PhysRevX.6.011029,PhysRevB.102.245105,PhysRevB.102.235105,PhysRevB.104.155142,PhysRevB.104.035107,Herbut:2023xgz}. 
However, the finite-temperature behavior of such strongly-correlated Dirac materials due to the interplay of critical order-parameter and chiral fermion excitations remains an open question. 
In addition, in view of the finite, albeit small, temperatures in actual experimental setups, going beyond the study of quantum critical exponents and developing a better understanding of thermal effects seems mandatory.

Here, we investigate quantum critical behavior at zero and finite temperature in Dirac semimetals, which can be considered as minimal models for metallic quantum criticality.
To that end, we describe universal and non-universal aspects in a field-theoretical framework by employing  functional methods.
This allows us to calculate phase diagrams and scaling behavior of such relativistic Dirac semimetals.
More concretely, we do not only calculate critical exponents of the underlying zero-temperature quantum critical points, but we also map out the different finite-temperature phases and the crossovers or phase transitions between them for the $(2+1)$-dimensional chiral Ising, chiral XY, and chiral Heisenberg models.
We estimate the extent of a quantum critical fan via the deviation from scaling with the dynamic critical exponent.
We calculate the quasiparticle weight of the Dirac fermions as an estimate for non-Dirac liquid behavior in analogy to non-Fermi liquids in metals. 
We show the phenomenon of precondensation, a strongly-fluctuating semimetallic regime, which arises as a precursor for the formation of order. We show how the Coleman--Hohenberg--Mermin--Wagner (CHMW) theorem is fulfilled within our approach and how signatures of the Berezinskii--Kosterlitz--Thouless (BKT) transition arise also in the chiral XY model.

\section{Models}
At low excitation energies the classical action for electrons in a Dirac material is represented by ${S=\int_0^{1/T}d\tau\int d^{d-1}x \mathcal{L}}$ with $\mathcal{L}=\mathcal{L}_0 + \mathcal{L}_\mathrm{int}$, 
where the non-interacting part $\mathcal{L}_0$ can be written as
\begin{equation}
\mathcal{L}_0[\bar{\psi},\psi]=  \bar{\psi} (I_{N_\mathrm{f}} \otimes\gamma^\mu) \partial_\mu \psi\,.
    \label{eq:NonIntAction}
\end{equation}
Here, ${\mu=0,1,...,d-1}$ stands for the spacetime index with $\partial_\mu$ being the $d$-dimensional spacetime derivative and we have set the Fermi velocity to unity.
The square matrices $\gamma^\mu$ obey the Clifford algebra, ${\{\gamma^\mu,\gamma^\nu\}=2\delta^{\mu\nu} I_{d_\gamma}}$, where $d_\gamma$ is the dimensionality of the spinors determined by the chosen algebra representation~\footnote{In the present work, the considered traces over $\gamma^\mu$ are independent of the specific chosen representation. We note that in perturbative calculations there can be differences at higher loop orders.}. 
We consider $N_\mathrm{f}$ fermion flavors yielding a ${(d_\gamma N_\mathrm{f})}$-component gapless Dirac field $\psi$ and its Dirac conjugate ${\bar{\psi}= \psi^\dagger (I_{N_\mathrm{f}} \otimes \gamma^0) }$, corresponding to, e.g, spin, layer, and/or (mini)-valley degrees of freedom.
For example, in the case of spin-1/2 electrons on graphene's honeycomb lattice one finds $N_\mathrm{f}=2$~\cite{Herbut2006} and charge neutral twisted bilayer graphene corresponds to $N_{\mathrm{f}}=4$~\cite{PhysRevLett.123.157601,Parthenios2023,huang2024angletunedgrossneveuquantumcriticality,biedermann2024twisttunedquantumcriticalitymoire}.

The Lagrangian part $\mathcal{L}_\mathrm{int}$ contains a sum of local quartic terms that describe interactions of electrons in a Dirac material, which are structurally given by~\cite{Herbut2009,Boyack:2020xpe}
\begin{equation}
    \mathcal{L}_\mathrm{int} [\bar{\psi},\psi]= \frac{g}{2}  \left(\bar{\psi} M_a \psi\right)\left(\bar{\psi} M_a \psi\right)\,. 
    \label{eq:InteractingAction}
\end{equation}
Here, we have introduced the dimensionful coupling constant $g$ and the index $a$ runs from 1 to $N$.
Together with $\mathcal{L}_0$, cf. Eq.~\eqref{eq:NonIntAction}, the full Lagrangian then corresponds to a Gross--Neveu- or Nambu--Jona-Lasinio-type four-fermion model~\cite{PhysRev.122.345,PhysRevD.10.3235,Zinn-Justin:1991ksq}.
The $(d_\gamma N_\mathrm{f})$-dimensional Hermitian matrices $M_a$ are chosen such that the interacting part, Eq.~\eqref{eq:InteractingAction}, satisfies $\mathrm{U}(1)$- and $\mathbb{Z}_2$-chiral, Lorentz, time-reversal and parity-inversion symmetries, just as Eq.~\eqref{eq:NonIntAction}~\footnote{Note that these symmetries forbid to have any second-order term or any quartic-order term ${\sim \left(\bar{\psi}  M_a \psi\right) \left(\bar{\psi}  M_b \psi\right)}$ with $M_a\neq M_b$, see Ref.~\cite{Herbut2009} for details.
In addition, the algebraic Fierz identities linearly constrain the number of quartic-order terms in the Lagrangian.}.

In the vicinity of a phase transition towards a specific kind of symmetry-breaking order, it is often convenient to introduce a Hubbard--Stratonovich field in that particular channel, e.g., ${h \phi_a \sim g \bar{\psi} M_a \psi}$, leading to the partially bosonized action
\begin{equation}
    \mathcal{L}_\mathrm{PB}[\bar{\psi},\psi,\phi_a] = \mathcal{L}_0[\bar{\psi},\psi] +   h  \phi_a \bar{\psi} M_a \psi +  
    \frac{1}{2} m_\phi^2 \phi_a \phi_a \,.
    \label{eq:HSaction}
\end{equation}
Generally, fluctuations will generate order-parameter dynamics and bosonic self-interactions. Hence, we add corresponding terms in the Hubbard--Stratonovich action~\eqref{eq:HSaction}, yielding a Gross--Neveu--Yukawa~(GNY) model
\begin{equation}
\begin{split}
    \mathcal{L}_\mathrm{GNY}&[\bar{\psi},\psi,\phi_a]=  \bar{\psi} ( I_{N_\mathrm{f}} \otimes\gamma^\mu) \partial_\mu \psi - \frac{1}{2}  \phi_a \partial_\mu^2 \phi_a  \\
    &+   h  \phi_a \bar{\psi} M_a \psi +   
    \frac{1}{2} m_\phi^2 \phi_a \phi_a +   
    \frac{1}{8} \lambda \left(\phi_a \phi_a\right)^2
    \,.
    \label{eq:GNYaction}
\end{split}
\end{equation}
This action describes Dirac excitations coupled via a Yukawa interaction, $h^2\equiv g m_\phi^2$, to an $N$-component (auxiliary Hubbard--Stratonovich) bosonic order-parameter field, $\phi=(\phi_1,...,\phi_N)$ with mass $m_\phi$, quartic-order coupling $\lambda$, and exhibits global $\mathrm{O}(N)$ symmetry~\cite{Herbut2006,PhysRevB.97.041117,Boyack:2020xpe,Herbut:2023xgz}.

As a concrete example, consider ${d_\gamma=4}$, which corresponds to the honeycomb lattice with its valley and sublattice degrees of freedom.
A possible choice for the ($4\times4$) gamma matrices in $d=2+1$ dimensions, describing spinless electrons, is given by ${\gamma^0=I_2\otimes\sigma_3}$, ${\gamma^1=\sigma_3\otimes\sigma_2}$ and ${\gamma^2=I_2\otimes\sigma_1}$ for the kinetic term, along with two additional anticommuting matrices ${\gamma^3=\sigma_1\otimes\sigma_2}$ and ${\gamma^5=\sigma_2\otimes\sigma_2}$, which together define the ``graphene representation''~\cite{Herbut2006}.

Including the spin degree of freedom amounts to choosing $N_\mathrm{f}=2$ in this representation. 
Then, the 8-dimensional square matrix ${M_a=\sigma_a \otimes I_{4}}$ with $a=0$ defines the chiral Ising model, where $\sigma_0\equiv I_2$. An index $a$ running from 1 to 3, with $\sigma_a $ being the Pauli matrices, defines the chiral Heisenberg model. 
These two models then effectively describe spin-1/2 fermions on the honeycomb lattice in the vicinity of the phase transition from a Dirac semimetal phase to charge-density wave~(CDW) or  antiferromagnetic~(AFM) order, respectively~\cite{Herbut2006}.

It was argued in Ref.~\cite{PhysRevResearch.5.043173} that the experimentally observed relativistic Mott transition in artificial graphene engineered from twisted WSe${}_2$ tetralayers~\cite{ma2024relativisticmotttransitionstrongly} simulates the ideal honeycomb lattice with sublattice and SU(2) spin rotation invariance where interactions drive an AFM transition. 
This suggests a transition described by the $N_{\mathrm f}=2$ chiral Heisenberg model.

One version of the chiral XY model corresponds to choosing ${M_a= I_{N_{\mathrm{f}}} \otimes \sigma_a \otimes \sigma_{2}}$ with $a=1,2$ 
and the bosonic field being a complex scalar field, representing a Kekul\'e distortion of spin-1/2 fermions on the honeycomb lattice for $N_{\mathrm{f}}=2$~\cite{PhysRevLett.98.186809}.
For $N_\mathrm{f}=4$, the chiral XY model was argued to be the effective model that describes the quantum phase transition to an IVC state in angle-tuned twisted bilayer graphene~\cite{huang2024angletunedgrossneveuquantumcriticality,biedermann2024twisttunedquantumcriticalitymoire}.

\section{Functional renormalization group}
\subsection{Method}
A suitable method to simultaneously resolve the models' universal critical behavior and non-universal phase diagrams near the quantum phase transition is the functional renormalization group (FRG)~\cite{Dupuis:2020fhh}.
The FRG includes fluctuations in a non-perturbative way and has been found to describe the  critical behavior of many models of statistical physics and quantum materials to a very good level of precision even at low orders of the approximation, see, e.g., Refs.~\cite{Litim:2001fd,PhysRevD.64.105007,PhysRevB.89.205403,PhysRevE.101.042113}.
It was employed before to directly access the quantum critical behavior of Dirac materials in $2+1$ spacetime dimensions at zero temperature, see, e.g., Refs.~\cite{PhysRevLett.86.958,PhysRevB.66.205111,PhysRevB.89.205403,PhysRevB.93.125119,Knorr2016,Classen2017,Knorr2018,PhysRevB.97.125137,PhysRevResearch.2.013034,PhysRevResearch.2.022005}, and 
to describe finite temperature effects in such models~\cite{SCHAEFER2005479,Braun_2012,Scherer:2013many,PhysRevD.90.076002,
stoll2021bosonicfluctuations1}.
The method is based on introducing an infrared (IR) cutoff with scale $k$ into the partition function, ${\mathcal{Z}=\int_\Lambda \mathcal{D}\Phi e^{-S\left[\Phi\right]}\to \mathcal Z_k}$, where $\Lambda$ is the ultraviolet~(UV) cutoff.
This is achieved by adding a regulator term bilinear in the fields ${\Phi(q)=\begin{pmatrix}
    \phi(q), 
    \psi(q), 
    \bar{\psi}^T(-q)
\end{pmatrix}^T}$ to the microscopic action, i.e.,
\begin{equation}
\begin{split}
    S\rightarrow S+\int\frac{d^{d}q}{(2\pi)^{d}}\int\frac{d^{d}p}{(2\pi)^{d}} &\Big[\frac{1}{2}\phi(-q) R_k^{(B)}(q,p)\phi(p)\\
    &+\bar\psi(q) R_k^{(F)}(q,p)\psi(p)\Big],
\end{split}
\end{equation}
with $R_k^{(B)}(q,p)$ and $R_k^{(F)}(q,p)$ being the bosonic and fermionic infrared regulators, respectively.
The scale-dependent flowing action $\Gamma_k$ is the Legendre transform of the scale-dependent Schwinger functional ${W_k=\ln \mathcal Z_k}$~\cite{Dupuis:2020fhh,Wipf2021}.

The FRG evolution of $\Gamma_k$ 
interpolates between the microscopic action at large scale, $k\rightarrow \Lambda$, where the classical action is recovered $\Gamma_{k\rightarrow\Lambda} \simeq S $, and the full effective action at $k\rightarrow 0$, i.e., $\Gamma_{k\rightarrow0} \simeq \Gamma $.
The evolution equation for the flowing action $\Gamma_k$ is  the Wetterich equation~\cite{Wetterich:1992yh}
\begin{equation}
    \partial_t \Gamma_k [\Phi] = \frac{1}{2} \text{STr} \left\{ \left[\Gamma_k^{(2)} [\Phi] + R_k\right]^{-1} \left(\partial_t R_k\right) \right\},
    \label{eq:WetterichEq}
\end{equation}
where the RG time is $t=\ln(k/\Lambda)$, with $\partial_t = k \frac{d}{dk}$. The second-order functional derivative of the effective action with respect to the fluctuating fields is given by
$
{\big(\Gamma_k^{(2)} \big)^{ab} (p,q) \equiv \frac{\overrightarrow{\delta}}{\delta \Phi_a^T(-p)} \Gamma_k \frac{\overleftarrow{\delta}}{\delta \Phi_b(q)}}\,,$
and the supertrace, $\mathrm{STr}$, integrates over all momenta and sums over all field degrees of freedom including a minus sign for the fermion sector.

At finite temperature, frequency integrals are replaced by Matsubara sums,
\begin{equation}
    q_0\to \omega_n = 2\pi c_n T,\quad \int\frac{d^dq}{(2\pi)^d}\to   T\sum_{n\in\mathbb{Z}}\int\frac{d^{d-1}q}{(2\pi)^{d-1}},
\end{equation}
where the Matsubara frequencies $\omega_n$ can be fermionic with the definition ${c_{n}\vert_\psi\equiv c_{F,n}=(n+1/2)}$, or bosonic with ${c_{n}\vert_\phi\equiv c_{B,n}=n}$.
The FRG approach has been well established for systems at finite-temperature in various contexts, before, e.g., for cold atoms, Hubbard models, or quantum chromodynamics, see~\cite{Dupuis:2021} for references.

\subsection{Flow equations}
To use the Wetterich equation in the context of the GNY models, we choose a truncation for the flowing action $\Gamma_k=\Gamma_k[\bar{\psi},\psi,\phi_a]$ reading
\begin{equation}
\begin{split}
\Gamma_k =& \int_0^{1/T}d\tau\int d^{d-1}x \Big[  Z_{\psi,k} \bar{\psi} \left( I_{N_\mathrm{f}} \otimes \gamma^\mu \right) \partial_\mu \psi \\
&- \frac{1}{2} Z_{\phi,k} \phi_a \partial_\mu^2 \phi_a + h_k   \phi_a \bar{\psi} M_a \psi + U_k(\rho)\Big]\,.
\label{eq:EffAction}
\end{split}
\end{equation}
This approach is well-established in the FRG literature and has been successfully applied to a wide range of systems, where even leading-order derivative expansions yield quantitatively accurate results near criticality as showed in Table~\ref{table:CriticalExponents}, see e.g.~\cite{Berges:2002,PhysRevB.89.205403}.
Here, we included scale-dependent wave function renormalizations $Z_{\phi,k}$ and $Z_{\psi,k}$ for bosons and fermions, respectively, a scale-dependent Yukawa coupling $h_k$, and a scale-dependent effective potential $U_k$, which depends on the field invariant  ${\rho=\frac{1}{2}\phi_a \phi_a}$. 
At the UV limit, the classical microscopic action given in Eq.~\eqref{eq:HSaction} is recovered by ${Z_{\phi,k\rightarrow \Lambda} \rightarrow 0}$ and ${Z_{\psi,k\rightarrow \Lambda} = 1}$.
The anomalous dimensions for fermions and bosons can be defined via the wave function renormalizations ${\eta_{\phi/\psi} = - \partial_t \ln Z_{\phi/\psi,k}}$.
We discuss the UV conditions of the effective potential and Yukawa coupling below.

The FRG flow equation for the dimensionless effective version of the effective potential, i.e., ${u_k=k^{-d}U_k}$, is obtained by evaluating Eq.~\eqref{eq:WetterichEq} at constant bosonic field and vanishing fermionic fields. To that end, we choose a linear regulator scheme~\cite{Litim:2001fd,PhysRevD.64.105007}, which is explicitly given in Appendix~\ref{app:ThresholdFunctions}.
We find
\begin{equation}
    \begin{split}
\partial_t u_k =& -d u_k + (d-2+\eta_\phi) \bar\rho u_k'\\
& - 2v_d d_\gamma N_\mathrm{f} l_0^{(F) d} (\tau, \omega_\psi; \eta_\psi)\\
&+ 2v_d l_0^{(B)d}(\tau, \omega_\phi ;\eta_\phi) \\
&+ 2(N-1)v_dl_0^{(B)d}(\tau,u_k';\eta_\phi),
\label{eq:PotentialFlow}
\end{split}
\end{equation}
where we introduced the fermionic and bosonic masses ${\omega_\psi=2 \bar h^2_k \bar\rho}$ and ${\omega_\phi=u_k'+2\bar\rho u_k''}$, respectively.
The bar notation indicates dimensionless rescaled quantities such as ${\bar\rho=  Z_{\phi,k} k^{2-d} \rho}$ and ${\bar h^2_k = Z_{\phi,k}^{-1} Z_{\psi,k}^{-2} k^{d-4} h_k^2}$,
while the prime notation indicates derivatives with respect to $\bar\rho$. The reduced temperature is ${\tau = 2\pi T/k}$ and the volume factor ${v_d^{-1}= 2^{d+1} \pi^{d/2} \Gamma(d/2)}$. 
The threshold functions $l^{(B/F)d}_0$ represent all one-loop corrections using the full fermion and boson propagators. 
They are a product of a factor coming from momentum integration and another one related to Matsubara summation, which encodes the systems dependence on the temperature. 
For our regulator, these integrations/summations can be carried out analytically, see Appendix~\ref{app:ThresholdFunctions}.

The first line in Eq.~\eqref{eq:PotentialFlow} is due to dimensional and field rescalings, the second line arises from the fluctuations of the Dirac fermions, the third line represents the  contributions from the radial modes of the bosons and the last line corresponds to the possible Goldstone-mode fluctuations, which are only present for the chiral XY ($N=2$) and chiral Heisenberg ($N=3$) models, but not for the chiral Ising case $N=1$. 
Schematically, fermion fluctuations drive the effective potential into the regime with spontaneously broken symmetry upon integration towards the infrared, i.e., $k\to0$. 
Bosonic fluctuations, in contrast, counteract this tendency and, in particular, the Goldstone fluctuations melt down a finite vacuum expectation value (VEV) at finite temperatures, i.e., they support the restoration of symmetry, see below.

We expand the effective potential as
\begin{equation}
    U_k(\rho) = \sum_{n=1}^\infty \frac{\lambda_k^{(n)}}{n!} (\rho-\kappa_k)^n,
    \label{eq:EffPotential}
\end{equation}
with the $k$-dependent minimum of the effective potential $\kappa_k=\rho_{\mathrm{min},k}$, related to the VEV of the bosonic field $\kappa_{k\to 0}=\frac{1}{2} \langle\phi_{a}\phi_{a}\rangle$. 
The effective action $\Gamma_k$ with this kind of potential and uniform wave function renormalization is referred to as improved local potential approximation~(LPA${}'$).

A finite VEV of the bosonic field is directly related to the fermionic mass gap in the IR limit ${m_\mathrm{f} \equiv h_k \sqrt{2\kappa_k} \rvert_{k\rightarrow 0}}$. 
Therefore, the VEV serves as an order parameter for spontaneous symmetry breaking (SSB), which renders the fermions massive.
We work with a basic truncation taking $n$ up to $n_{\mathrm{max}}=2$ in Eq.~\eqref{eq:EffPotential}, which we find to be sufficient for our purposes, see below.
We use the notation ${\lambda_k^{(1)}\equiv m_{\phi,k}^2}$ for the dimensionful bosonic mass squared in the symmetric (SYM) regime (${\kappa_k=0}$), ${\lambda_k^{(1)}\equiv 0}$ in the SSB regime (${\kappa_k\neq0}$), and ${\lambda_k^{(2)}\equiv\lambda_k}$ being the dimensionful quartic  
bosonic coupling.

We note that we use a truncation with an effective Lorentz invariance due to equal velocities of the bosons and the fermions, i.e.,  $v_\mathrm{b}=v_\mathrm{f}$, which we have set to unity. 
This invariance was shown to emerge at the QCP of Gross--Neveu-like models, see Ref.~\cite{Roy:2015zna}, and we assume that remains a good approximation in proximity to the QCP.
This assumption is further supported by continuity arguments, suggesting that the system remains approximately Lorentz-invariant close to the QCP, even at finite temperature. 
Additionally, recent Quantum Monte Carlo results \cite{PhysRevB.110.125123} provide numerical evidence for this emergent symmetry in Dirac systems near criticality.

The flow equations for the Yukawa couplings and for the fermionic and bosonic wave function renormalizations are obtained from the following projection prescriptions 
\begin{equation}
\begin{split}
    h_k = \frac{1}{  N_\mathrm{f} d_\gamma} \mathrm{Tr}\left[ M_a \frac{\overrightarrow{\delta}}{\delta \Delta \phi_a(q)} \frac{ \overrightarrow{\delta}}{\delta \bar\psi(q^\prime)} \Gamma_k \frac{ \overleftarrow{\delta}}{\delta \psi(q^{\prime\prime})} \right] ,
    \label{eq:PRYukawaAll}
\end{split}
\end{equation}
\begin{equation}
\begin{split}
    Z_{\phi,k} = \frac{\partial}{\partial q^2} \int \frac{d^d q^\prime}{(2\pi)^d} \frac{\overrightarrow{\delta}}{\delta \Delta\phi_a(-q)}  \Gamma_k \frac{\overleftarrow{\delta}}{\delta \Delta\phi_a(q^\prime)}\, ,
    \label{eq:bAnomalousDimensionPrescription}
\end{split}
\end{equation}
and
   \begin{equation}
\begin{split}
   &Z_{\psi,k} = -\frac{i}{ N_\mathrm{f} d d_\gamma} \\
   &\,\times\mathrm{Tr}\left[(I_{N_\mathrm{f}}\otimes\gamma^\mu) \frac{\partial}{\partial q^\mu} \int \frac{d^d q^\prime}{(2\pi)^d} \frac{\overrightarrow{\delta}}{\delta \bar{\psi}(-q)} \Gamma_k\frac{\overleftarrow{\delta}}{\delta \psi(q^\prime)} \right],
    \label{eq:fAnomalousDimensionPrescription}
\end{split}
\end{equation}
by substituting Eq.~\eqref{eq:WetterichEq} in their scale derivatives. 
All the expressions are evaluated at $\bar{\psi}=\psi=\Delta\phi_a=0$ and zero momenta $q=q^\prime=q^{\prime\prime}=0 $.
In Appendix~\ref{app:Projections}, we give the flow equations and their derivation explicitly for the Yukawa couplings and the anomalous dimensions, which are different for the chiral Ising, 
chiral XY, and 
chiral Heisenberg model. 
During the FRG evolution, the flow equations of the boson and fermion anomalous dimensions as well as the Yukawa coupling are then evaluated at the scale-dependent minimum of the effective potential.

\section{Quantum critical exponents}

Quantum critical points show a specific  behavior that can, for example, be represented by the scaling of the correlation length at zero temperature, i.e., it diverges as
\begin{align}
    \xi\propto |g-g_c|^{-\nu}\,,
\end{align}
where $\nu$ is the correlation length exponent.
Simultaneously, the characteristic energy gap $\Delta$ scales as
\begin{align}
    \Delta \propto |g-g_c|^{z\nu}\,,
\end{align}
with the dynamical critical exponent $z$.
Importantly, the quantum critical point has a significant impact on the phase diagram even at finite temperature, e.g., by governing its behavior within the quantum critical fan (QCF) above the QCP.
For example, the thermal length in the QCF then scales as
\begin{align}
    \xi_T\propto T^{-1/z}\,.
\end{align}
Further relevant relations on quantum critical behavior at zero and finite temperature can be found below and in standard reviews, e.g, Ref.~\cite{Vojta_2003} and textbooks, e.g., Ref.~\cite{Sachdev2011}.

We extract the critical exponents of the QCP from the properties of the non-Gaussian infrared-stable fixed point of the models' FRG equations~\cite{PhysRevLett.86.958,PhysRevB.66.205111,PhysRevB.89.205403}. 
To that end, we consider $T=0$ and determine the coupling coordinates where all FRG flow equations vanish simultaneously, i.e., ${\beta_{\alpha_i}(\alpha^*)=\partial_t \alpha_i \rvert_{\alpha=\alpha^*}=0}$ with ${\alpha^*=\big\{h_k^*, \lambda^{(n)*}_k\big\}}$.
The critical exponents then correspond to the negative eigenvalues $\theta_i$ of the stability matrix, $S_{ij} = \frac{\partial \beta_{\alpha_i}}{\partial \alpha_j}\big\rvert_{\alpha=\alpha^*}$ evaluated at the non-Gaussian fixed point $\alpha^*$.
We order the eigenvalues $\theta_i$ by magnitude and the largest positive one corresponds to the inverse correlation length exponent $\theta_1=1/\nu$ and the anomalous dimensions at the QCP are given by Eqs.~\eqref{eq:AnomalousDimPhiFlow} and~\eqref{eq:AnomalousDimPsiFlow} evaluated at $\alpha^\ast$.

\begin{table}[t!]
\centering
\begin{tabular}{ p{4cm} p{1.4cm} p{1.4cm} p{1.4cm}}
 \hline \hline
chiral Ising& $1/\nu$ & $\eta_\phi$ &$\eta_\psi$\\ 
 \hline
  \textit{this work}, FRG (LPA${}'$4) & 1.00  & 0.76 & 0.032 \\
  FRG (NLO) \cite{Knorr2016}  & 0.994(2) & 0.7765 & 0.0276 \\
  $\epsilon$-exp w/ DREG3 \cite{PhysRevB.98.125109}  & 0.993(27) & 0.704(15) & 0.043(12) \\
     conformal bootstrap \cite{Erramilli2023}  & 0.998(12) & 0.7329(27) & 0.04238(11) \\
     QMC \cite{PhysRevB.108.L121112}  & 1.07(12) & 0.72(6) & 0.04(2) \\
  \hline \hline
chiral XY& $1/\nu$ & $\eta_\phi$ &$\eta_\psi$\\ 
 \hline
  \textit{this work}, FRG (LPA${}'$4)  & 0.86  & 0.87 & 0.063  \\
  FRG (LPA${}'$12) \cite{Classen2017} & 0.86 & 0.88 &  0.062 \\
$\epsilon$-exp (interpolation)  \cite{Hawashin2025} & 0.904(9) & 0.85(3) & 0.095(19) \\
  QMC \cite{Otsuka:2018} & 0.94(1) & 0.64(2) &  0.15(1)\\
  \hline \hline
chiral Heisenberg & $1/\nu$ & $\eta_\phi$ &$\eta_\psi$\\ 
 \hline
 \textit{this work}, FRG (LPA${}'$4)  & 0.76 & 1.01 & 0.085  \\
 FRG (NLO) \cite{Knorr2018} & 0.795 & 1.032 & 0.071\\
$\epsilon$-exp (interpolation) \cite{PhysRevB.107.035151} & 0.83(12) & 1.01(6) & 0.13(3)   \\
HMC \cite{PhysRevB.104.155142} & 0.84(4) & 0.52(1) & \\
DQMC \cite{PhysRevB.104.035107} & 1.11(4) & 0.80(9) & 0.29(2)
\end{tabular}
\caption{\textbf{Chiral Ising/ XY/Heisenberg universality} in $2+1$ spacetime dimensions for $N_{\mathrm{f}}=2$: correlation length exponent $\nu$ and anomalous dimensions $\eta_\phi$ and $\eta_\psi$ for bosons and fermions, respectively.
We compare to results from more advanced FRG truncation schemes, including both next-to-leading-order (NLO) approximations and higher-order derivative expansions such as LPA${}'$12.
We also compare our results with estimates from other methods, e.g., perturbative RG, conformal bootstrap, and Monte Carlo simulations.
To obtain the values for $\nu$ from some of the references, we have performed simple numerical inversion of $1/\nu$.}
\label{table:CriticalExponents}
\end{table}

For $N_{\mathrm{f}}=2$ and  $n_{\mathrm{max}}=2$, we show our estimates for the correlation length exponent $\nu$
and the anomalous dimensions $\eta_\phi$ and $\eta_\psi$ in Tab.~\ref{table:CriticalExponents}. 
Despite the simplicity of the present FRG truncation these estimates agree well with the results from more advanced FRG truncations schemes, see Tab.~\ref{table:CriticalExponents}.
For comparison, we also show the estimates from
other quantum many-body methods, e.g., perturbative RG, conformal bootstrap, or quantum Monte Carlo simulations in Tab.~\ref{table:CriticalExponents}. 
We note that especially for the case of the chiral Heisenberg model the estimates still vary quite strongly, which deserves further study.
For larger $N_{\mathrm{f}}$, for example, for the physically relevant choice of $N_{\mathrm{f}}=4$, the agreement improves systematically due to the approach of the large-$N_{\mathrm{f}}$ regime, see Refs.~\cite{Zerf2017,Erramilli2023,PhysRevB.107.035151,Hawashin2025} for the actual numbers.

\section{Finite-temperature phase diagram}

The analysis of the critical exponents provides a foundation for studying the phase diagram at finite temperature. Specifically, we are interested in the phases and excitations as function of the UV initial quartic fermion coupling $g=h_k^2/m_{\phi,k}^2$, which we obtain from the solution of the RG flow equations in the IR at $k\rightarrow0$.
To that end, we resolve the scale-dependence of the models' couplings, i.e., the effective potential, the Yukawa coupling, and the anomalous dimensions. 
For the UV initial conditions, we set the reference scale to $\Lambda=1$, defining the units.
We then choose an effective potential in the SYM regime, i.e., a vanishing VEV of the bosonic field and ${m_\phi^2|_{k=\Lambda}>0}$, and further we set ${\lambda|_{k=\Lambda}=0}$, ${h^2|_{k=\Lambda}=1}$, ${Z_\phi|_{k=\Lambda}=10^{-6}}$ and ${Z_\psi|_{k=\Lambda}=1}$, which reflects that the models are derived from a purely fermionic system.
Generally, the non-universal features of the phase diagram depend on the choice of these initial conditions and, for future work, it will be interesting to connect them to relevant microscopic lattice models. 
We have numerically confirmed that also the non-universal features of the phase diagram depend on the initial conditions only mildly.

\begin{figure}
    \centering
\includegraphics[width=\linewidth]{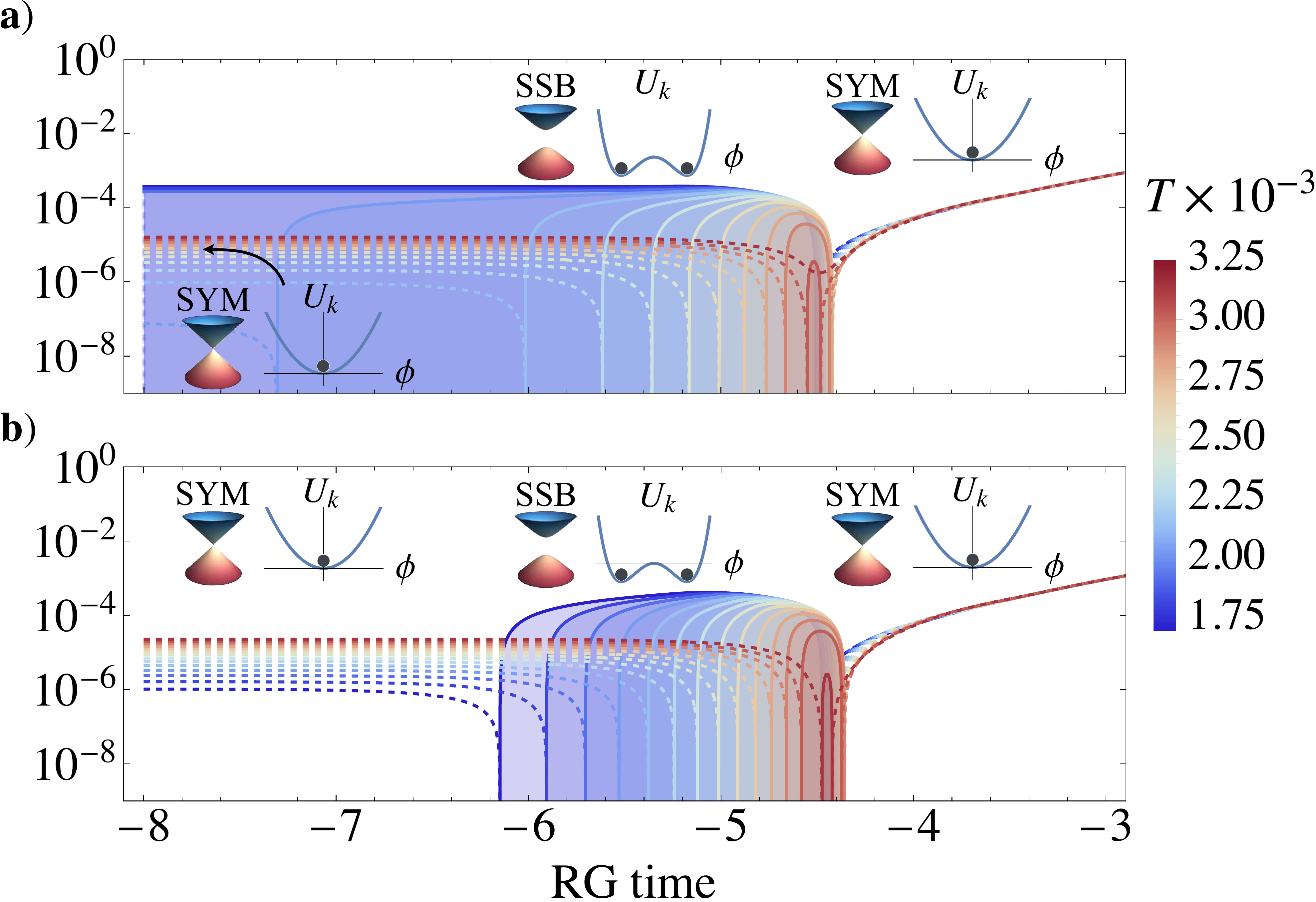}
\caption{\textbf{RG flows of effective potential} for the chiral Ising (top panel, $g=4.8125$) and the chiral Heisenberg model (bottom panel, $g=4.515$) at different $T$.
We exhibit flows starting from the UV scale $\Lambda=1$ with a potential in the SYM regime, represented by dashed lines for the parameter $m_{\phi,k}^2$.
\textbf{a)}~For the chiral Ising model, the $\mathbb{Z}_2$ symmetry can be broken at some RG time $t$ and develop a non-vanishing VEV (solid lines show $\kappa_k$). The VEV survives the IR limit at low but finite $T$ and gives rise to a gapped Dirac phase.
\textbf{b)}~In the chiral Heisenberg model, the fermion fluctuations drive the system into a symmetry-broken regime on intermediate scales (precondensation), but the condensate does not survive in the IR limit and the Dirac fermions remain gapless for all finite~$T$.
This is a manifestation of the CHMW theorem.
The chiral XY model also shows precondensation only on intermediate scales, but not condensation, see Sec.~\ref{sec:bkt}.
}
\label{fig:MW}
\end{figure}

\begin{figure*}
\centering
\includegraphics[width=\linewidth]{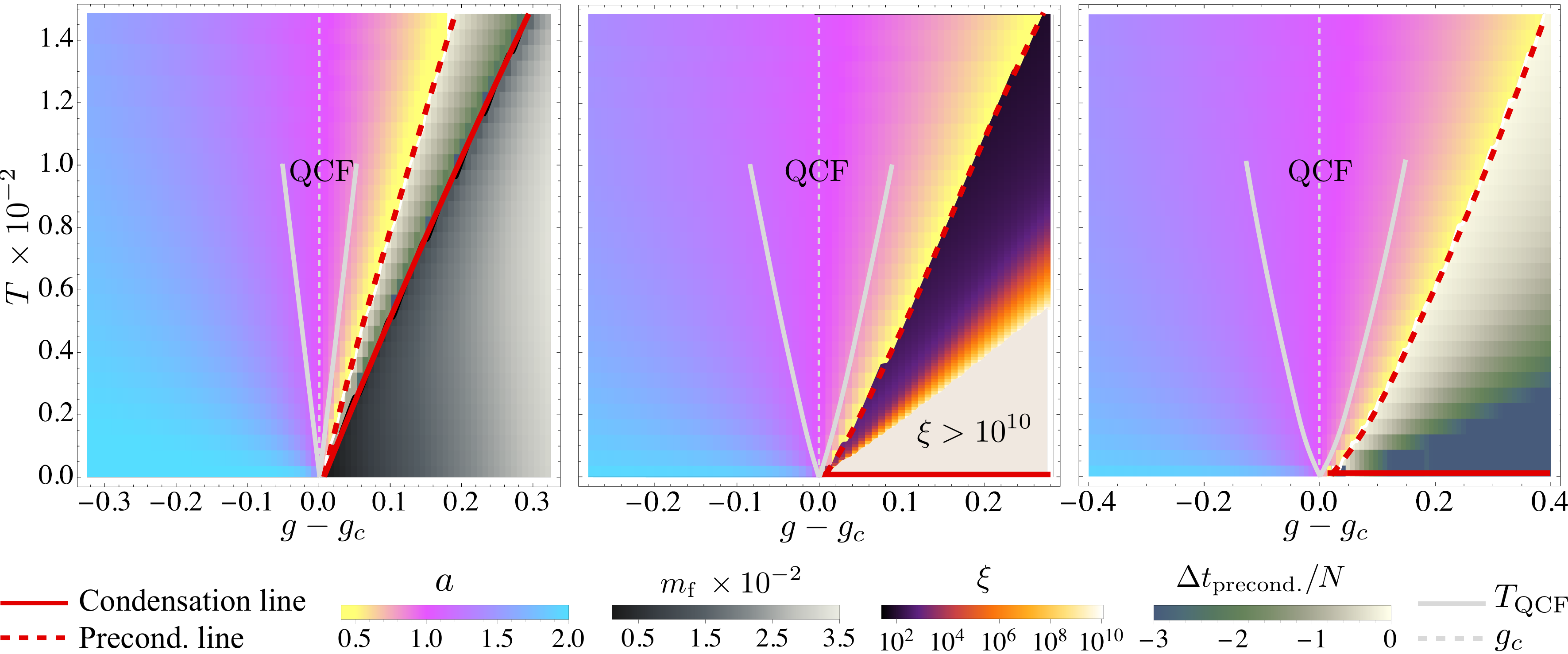}
\caption{
\textbf{Phase diagrams of the chiral Ising, chiral XY, and chiral Heisenberg models in the coupling-temperature plane} (from left to right) with $N_\mathrm{f}=2$. The red dashed lines identify the crossover to a semimetallic precondensation regime, while the red solid lines represent the actual second-order phase transition to a (Mott) insulator. 
The critical coupling at zero temperature is indicated with dashed gray vertical lines at $g_{c,\mathrm{I}}\simeq4.77$, $g_{c,\mathrm{XY}}\simeq4.62$ and $g_{c,\mathrm{H}}\simeq4.395$. 
In the symmetric regime of the phase diagram, we represent the exponent~$a$ of the thermal coherence length via a color scale.
The extent of the QCF is shown with solid light gray lines, which follow the power law given in Eq.~\eqref{eq:QCF} with $z=1\pm 0.2$ on top of the QCP.
For the chiral Ising model, we show the size of the fermionic mass gap, ${m_\mathrm{f}=\sqrt{2h^2\kappa}}$, in the symmetry-broken phase via the color code. Between the precondensation and the condensation lines, the color encodes the extent of the precondensation regime, $\Delta t_\mathrm{precond.}/N$, for the chiral Ising and chiral Heisenberg models.
For the chiral XY model, we show the correlation length in the precondensation regime.
The extremely fast growth of the correlation length was argued to indicate the transition into a BKT phase, 
see main text and Fig.~\ref{fig:BKTfit}.}
\label{fig:CriticalFan}
\end{figure*}

We illustrate the scale dependence of the effective potential in Fig.~\ref{fig:MW}, where we show the RG flow of $m_{\phi,k}^2$ (dashed lines) in the SYM regime starting in the UV for different temperatures. 
The effective potential 
evolves as the RG scale is decreased and, for small enough temperatures $T$, it develops a non-vanishing minimum $\kappa_k>0$ (solid lines), i.e., the system is in the SSB regime, where $\mathrm{O}(N)$ is broken.
When the system remains in the SSB regime in the IR limit, then the Dirac fermions receive a finite mass gap. 
Otherwise, when the SYM regime is restored by the bosons, we refer to this phenomenon as precondensation, which is discussed in more detail below.
In the upper panel of Fig.~\ref{fig:MW}, we show representative flows of the chiral Ising model for parameters near the QCP. 
Here, a non-vanishing VEV corresponds to the breaking of discrete $\mathbb{Z}_2$ symmetry, which is allowed by the CMWH theorem.
In contrast, the lower panel shows similar flows for the chiral Heisenberg model which exhibits a continuous O(3) symmetry that cannot be broken in two spatial dimensions at finite temperature, thereby leading our solution to always flow into the SYM regime in the IR in this case. 
A simple analytical reasoning for how this phenomenon is included in the FRG 
shows that boson fluctuations drive the unrenormalized VEV
to zero at $T>0$ for $d-1\leq 2$ and $N>1$~\cite{PhysRevLett.134.041602}. 
From the behavior of the solution in the IR, we determine the condensation transition to the insulating (Mott) phase as depicted in the phase diagram~Fig.~\ref{fig:CriticalFan}. 

As a next step, we explore the temperature dependence of the thermal length scale $\xi_T = 1/m_\phi|_{k\to0}$ of the system as a function of the initial coupling $g$.
We focus on the region of the phase diagrams where no condensation or precondensation occurs, i.e., in the fully symmetric phase where the Dirac fermions are gapless.
For fixed coupling~$g$, the thermal length exhibits a temperature dependence for which we choose the following power-law ansatz
\begin{equation}
    \xi_T \propto T^{-1/a},
    \label{eq:LengthScale}
\end{equation}
with an exponent $a$, which is to be determined.
We note that this ansatz is only valid within respective scaling regions, which can be found by numerically evaluating~$a$ for different temperatures and couplings, see Fig.~\ref{fig:CriticalFan}.

In Fig.~\ref{fig:CriticalFan}, we find two main scaling regions for the thermal length scale in the symmetric regime. 
The first one corresponds to the quantum critical fan above the QCP, where the exponent $a$ matches the dynamic critical exponent $z$ (pink regions), which follows directly from the definition of the dynamic exponent~\cite{Sachdev2011,Vojta_2003}. 
Note that $z=1$ at the QCP due to Lorentz invariance.
In addition, our calculation provides an estimate for the actual extent of this scaling region, see also the more quantitative analysis in the next paragraph.
The second scaling region is in the regime of smaller couplings, $g<g_c$, at small $T$. 
Here the exponent takes the value $a=2$ (cyan regions), which arises as a consequence of dominating Dirac-fermion fluctuations, see Appendix~\ref{app:a2} for a detailed reasoning. 
We note that a third scaling region is expected in the chiral Ising model due to the classical critical behavior near the finite-temperature $\mathbb{Z}_2$ phase transition.
This behavior, however, cannot be accessed within our truncation in two-spatial dimensions as all bosonic operators $\lambda^{(n)}_k\rho^{n}$ in Eq.~\eqref{eq:EffPotential} are canonically relevant.
A full resolution of the effective potential beyond a Taylor expansion can resolve this scaling at the phase transition within the FRG approach presented here~\cite{Dupuis:2020fhh,PhysRevLett.110.141601,PhysRevE.101.042113} which is, however, outside the focus of the present work.

\subsection{Quantum critical fan}

In the region of the quantum critical fan (QFC), the thermal length scale follows a power-law temperature dependence as in Eq.~\eqref{eq:LengthScale} with $a=z$ \cite{Torroba2020}. 
The theory of quantum phase transitions predicts the boundaries of this scaling region to be given by~\cite{Sachdev2011,Vojta_2003}
\begin{align}
    T_\mathrm{QCF} = A_\pm|g-g_c|^{z\nu}\,,
    \label{eq:QCF}
\end{align}
with the corresponding correlation length exponent $\nu$, see Tab.~\ref{table:CriticalExponents}.
The transition away from this quantum critical scaling region is not sharp, but rather a smooth crossover.
The prefactors $A_\pm$ are non-universal, and they can generally be different on the two sides of the QCP. 

We numerically determine $A_\pm$ from fitting Eq.~\eqref{eq:QCF} to our data in Fig.~\ref{fig:CriticalFan} based on a $\pm 20\%$~deviation of the dynamic critical exponent~$z$ from its QCP value $z=1$.
For our models with the given initial conditions, we find for the chiral Ising model that ${A_{+,\mathrm{Ising}} \simeq 0.190}$ and ${A_{-,\mathrm{Ising}}\simeq 0.192}$, for the chiral XY model that ${A_{+,\mathrm{XY}} \simeq 0.169}$ and ${A_{-,\mathrm{XY}}\simeq 0.176}$, and for the chiral Heisenberg model that ${A_{+,\mathrm{Heisenberg}} \simeq 0.122}$ and ${A_{-,\mathrm{Heisenberg}}\simeq 0.151}$, see the gray solid lines in Fig.~\ref{fig:CriticalFan}.
Note that due to Eq.~\eqref{eq:QCF} for the chiral Ising model, one therefore expects a cusplike QCF, where the boundary is a straight line as $z\nu\approx 1$; for the chiral XY model with $z\nu\approx 1.16$ and the chiral Heisenberg model with $z\nu\approx 1.31$, instead, the fan opens with increasing slope away from the QCP as shown in Fig.~\ref{fig:CriticalFan}.
For more quantitative details on the QCF, see Appendix~\ref{app:YukawaCouplings}.

\subsection{Precondensation regime}

Precondensation is a precursor to the formation of order, where the system has correlated domains of finite size $\sim k_{\mathrm{precond.}}^{-d}$. 
This phenomenon can be considered analogous to the formation of finite magnetic domains known from the ferromagnetic transition slightly above the Curie temperature.

The precondensation regime is characterized by strong order-parameter fluctuations, which make the fermions less coherent close to the QCP. For example, this manifests via a strongly reduced quasiparticle weight (see Sec.~\ref{sec:qpw}). 
Roughly speaking, the precondensation appears between the mean-field and the actual transition temperature. 
In the RG picture, we can identify regions of precondensation where condensation appears at intermediate RG scales but does not persist in the infrared limit. 
These RG scales are directly related to the finite size of correlated domains.
More concretely, the condensate vanishes, again, at some finite scale ${k_{\mathrm{precond.}}>0}$, i.e., there is a finite range of (logarithmic) scales $\Delta t_{\mathrm{precond.}}$ where the system is in the symmetry-broken regime. Eventually, ${k_{\mathrm{precond.}}\to 0}$ or ${\Delta t_{\mathrm{precond.}}\to -\infty}$ when approaching the true critical temperature, see Fig.~\ref{fig:CriticalFan}.
To identify the crossover line to the precondensation regime in Fig.~\ref{fig:CriticalFan}, we determine the temperature $T$ for a fixed coupling $g$ where a condensate first develops and then melts again during the flow, as shown in Fig.~\ref{fig:MW}.
This phenomenon also occurs in phase diagrams of ultracold atoms~\cite{Boettcher_2012} and models for quantum chromodynamics~\cite{khan2015phasediagramqc2dfunctional}. 

In the chiral XY and chiral Heisenberg models, condensation is restricted to zero temperature, as it cannot survive finite temperatures due the CHMW theorem.
As explained before, we can observe this in the potential flows in Fig.~\ref{fig:MW}. 
In the chiral Ising model discrete $\mathbb{Z}_2$ SSB occurs at both, zero and finite temperatures. 
In the chiral XY and chiral Heisenberg models, instead, the continuous chiral ${\mathrm{U}(1)\cong \mathrm{O}(2)}$ and spin ${\mathrm{SU(2)}\cong \mathrm{O}(3)}$ 
symmetries as well as the discrete $\mathbb{Z}_2$ symmetry are spontaneously broken only at zero temperature \cite{PhysRevB.97.125137,PhysRevB.89.205403}, see the precondensation and condensation transition lines in Fig.~\ref{fig:CriticalFan}. This leads to a particularly large precondensation regime in the chiral XY and chiral Heisenberg models.

\begin{figure}[t!]
    \centering
    \includegraphics[width=\linewidth]{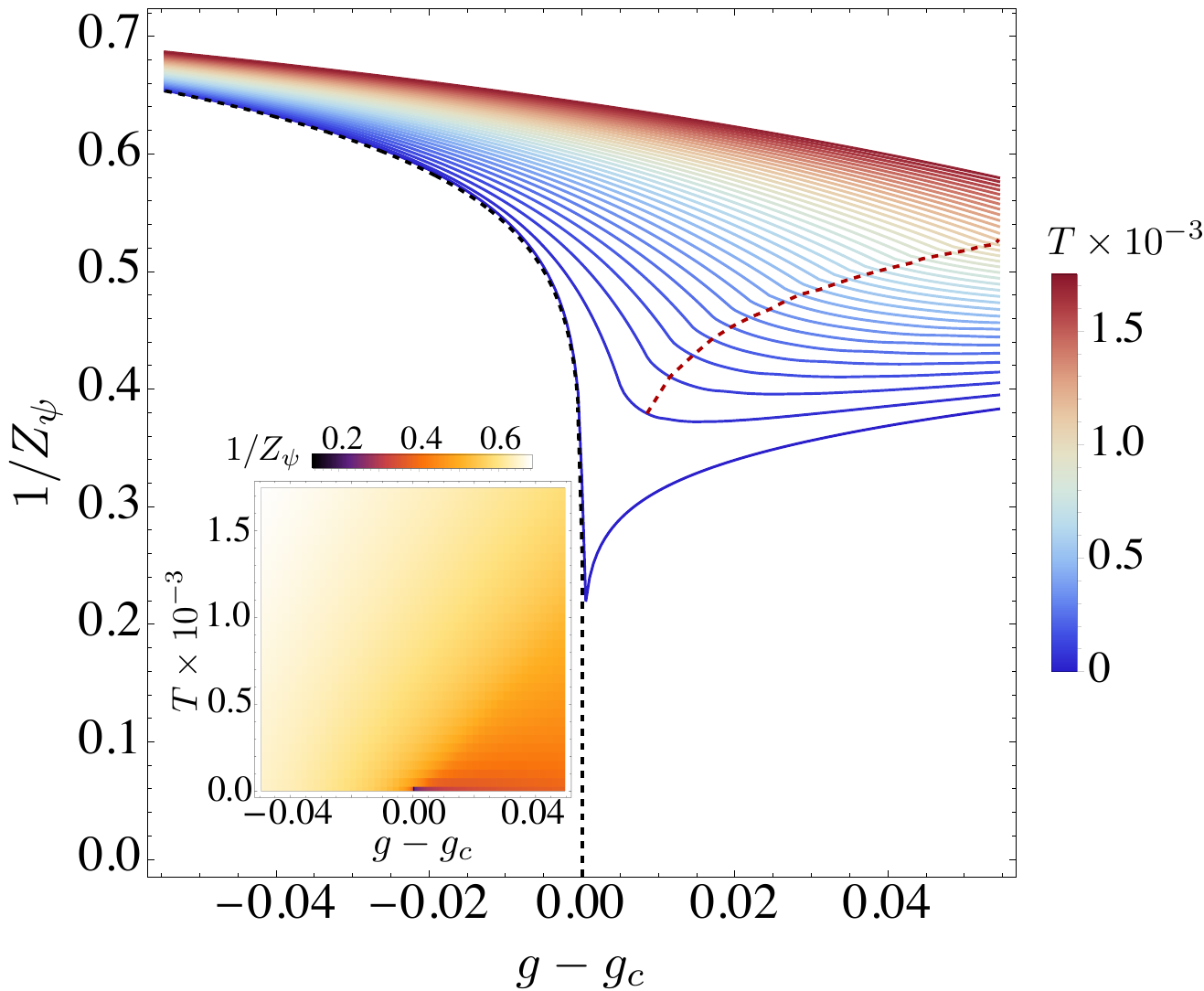}
    \caption{\textbf{Fermionic quasiparticle weight} given by the inverse of the fermionic wave function renormalization $1/Z_{\psi,k}$ as a function of the ultraviolet coupling $g$ and temperature $T$ for the chiral Heisenberg model with $N_\mathrm{f}=2$. The black dashed line corresponds to the fit provided in Eq.~\eqref{eq:HerbutFit}.
    The red dashed line corresponds to the precondensation line.
    Inset: fermionic quasiparticle weight as function of temperature and distance to the critical point.}
    \label{fig:NDLH}
\end{figure}

\subsection{Quasiparticle weight}
\label{sec:qpw}

The inverse of the wave function renormalization, $Z_{\psi,k}$, gives an estimate for the fermionic quasiparticle weight.
Quantum critical behavior is observed for vanishing fermionic quasiparticle weight, which is an indicator for the absence of quasiparticles.
This phenomenon only occurs at very low temperatures close to the QCP, which is the only point where the wave function renormalization diverges. 
The exhibited behavior is similar to the one of a non-Fermi liquid with an underlying extended Fermi surface or line~\cite{Xu2017}. 
Here, instead, due to the point-like Dirac nodes, we observe indications for non-Dirac liquid behavior, where the bosonic fluctuations induce strong damping of the fermions, see Fig.~\ref{fig:NDLH}.
The quasiparticle weight close to the QCP scales as \cite{Herbut2009} 
\begin{equation}
    1/Z_{\psi,k} \sim \abs{g-g_c}^{\nu \eta_\psi}
    \label{eq:HerbutFit}
\end{equation}
with an expected peak at $g_c$, which matches our numerical results very well, see Fig.~\ref{fig:NDLH} and~Appendix~\ref{app:QpIXY}.

At finite temperatures, we also find a pronounced suppression which is asymmetric on both sides of the QCP (Fig~\ref{fig:NDLH}). 
At the lowest temperatures, the quasiparticle weight 
is minimal above the QCP. Interestingly, at slightly larger temperatures, the minimum evolves to an inflection point which follows the crossover to the precondensation regime and disappears further away from the QCP. 
This reflects the strong impact of boson fluctuations in this regime which tend to make the fermions less coherent.
We observe analogous behavior for the chiral Ising and chiral XY model, see Appendix~\ref{app:QpIXY}.

\subsection{Signatures of BKT physics}\label{sec:bkt}

The chiral XY model in $2+1$ dimensional systems is special at low but finite temperatures; while phase fluctuations prevent long-range order at finite temperature, quasi-long-range order is still allowed, giving rise to the Berezinskii--Kosterlitz--Thouless~(BKT) transition~\cite{Berezinsky:1970fr,Kosterlitz:1973xp}. 
The BKT transition is a topological phase transition characterized by algebraic decay of phase correlations of an XY order parameter below $T_\mathrm{BKT}$. 
One hallmark of BKT physics is that for temperatures above but near the transition, the system exhibits essential scaling in the correlation length given by 
\begin{equation}
    \xi \sim e^{c /\sqrt{T-T_\mathrm{BKT}}}\,.
    \label{eq:BKTscaling}
\end{equation}
Here $c$ is a system-dependent positive constant.
Another hallmark is that, within the low-temperature BKT phase, the phase stiffness $J$ is finite.

\begin{figure}[t!]
\centering
\includegraphics[width=\linewidth]{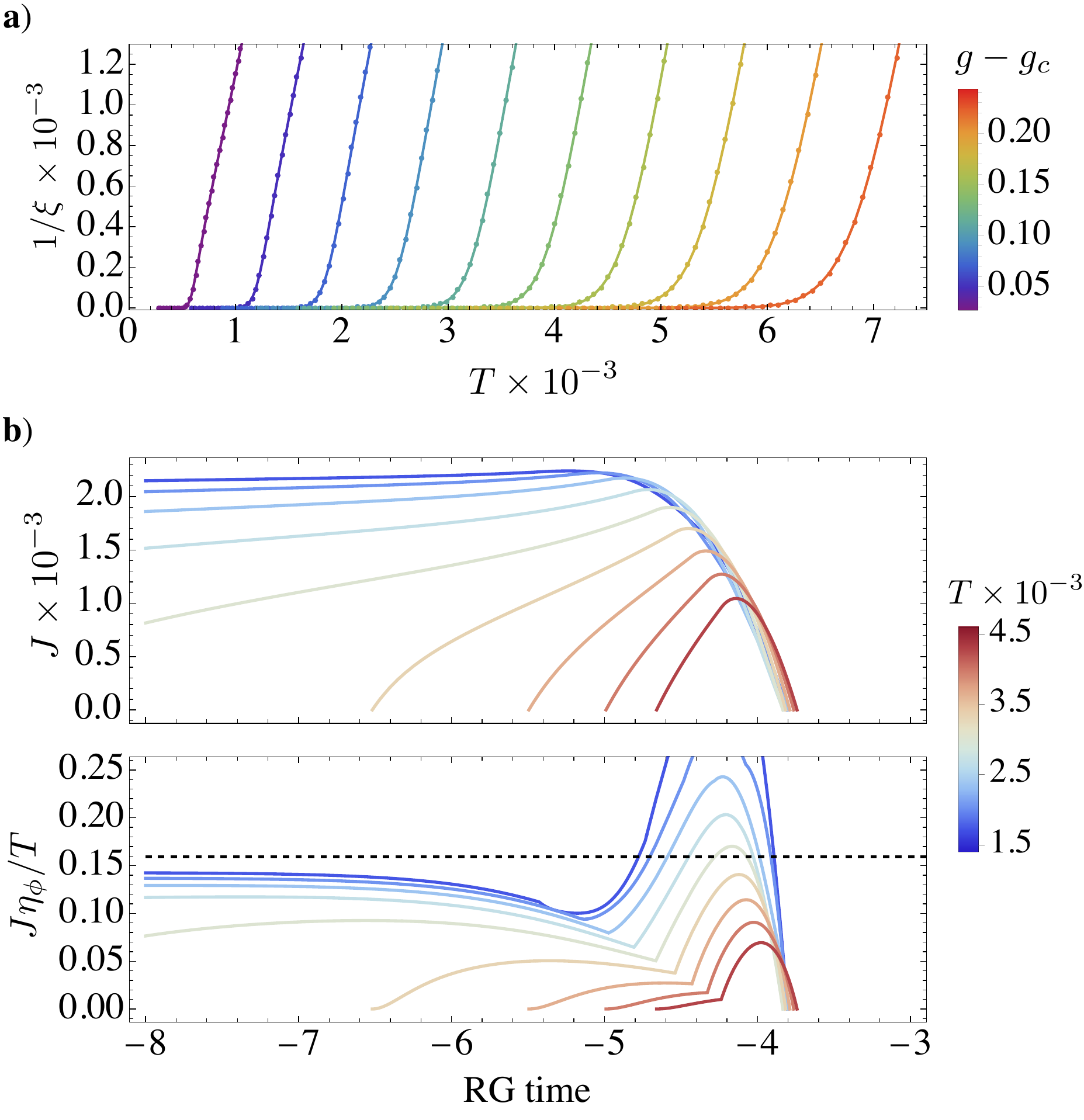}
\caption{ \textbf{Signatures of BKT behavior.} \textbf{a)} Evidence for approximate essential scaling of the correlation length as a function of temperature for different ultraviolet couplings following Eq.~\eqref{eq:BKTscaling} for the chiral XY model with $N_\mathrm{f} = 2$ near the QCP. 
\textbf{b)} RG flow of the phase stiffness $J$ and $J \eta_\phi/T$ for different temperatures $T$ at ultraviolet coupling $g=4.745$.
For low $T$, the flow of the phase stiffness is logarithmically slow and $J$ remains finite over many orders of magnitude, before it eventually deviates from approximate scaling within our approach.
The horizontal black dashed line in the lower panel represents the low-temperature limit $1/(2\pi)$.}
\label{fig:BKTfit}
\end{figure}

It has been argued in previous work on bosonic theories that an FRG approach can show signatures of BKT physics, already within the basic truncation for the bosonic action that we have employed in this work~\cite{PhysRevLett.75.378,PhysRevB.64.054513,PhysRevE.90.062105,Jakubczyk:2016rvr,Defenu2017}.
These signatures include that the inverse correlation length drops dramatically when approaching the BKT transition temperature from above, see Fig.~\ref{fig:BKTfit}. 
We note, however, that within the standard FRG truncations, the correlation length never actually diverges, see Refs.~\cite{Jakubczyk:2016rvr,Defenu2017} for a more detailed account of the FRG approach to BKT physics. 
Hence, essential scaling is only reproduced approximately and the correlation length stays finite, albeit huge.  Here, within a similar truncation scheme, we calculate the correlation length for the chiral XY model. 
The results are shown in the middle panel of Fig.~\ref{fig:CriticalFan} and in the top panel of Fig.~\ref{fig:BKTfit}.

In addition, we calculate the phase stiffness, which is given by the renormalized VEV  ${J=2\kappa_k}$. 
It exhibits a very long RG flow, i.e., it is approximately constant for low temperatures as expected for the BKT phase, see Fig.~\ref{fig:BKTfit} and Ref.~\cite{Jakubczyk:2016rvr} for comparison.
Eventually, however, it always vanishes in our approach.
Moreover, we find that the flow of the quantity $J\eta_\phi/T$  approaches the low-temperature limit $1/(2\pi)$, see Fig.~\ref{fig:BKTfit}.
Here, while not going beyond previous truncations in the FRG approach, we exhibit this evidence for BKT behavior for the first time in a system with gapless Dirac fermions near a transition corresponding to an $\mathrm{O}(2)$ order parameter.

\section{Conclusions and Outlook}

We presented a comprehensive approach that describes the phase diagram and (quantum) critical behavior in $2+1$ spacetime dimensions of chiral O($N$) models at zero and finite temperature, employing functional renormalization group methods.
We argued that our approach provides a good quantitative description of the quantum critical behavior at zero temperature in terms of the extracted critical exponents in agreement with previous estimates.
At finite temperature, our calculations expose the existence of a quantum critical fan above the quantum critical point and we determine the extent of this crossover region.
We also determine regions of precondensation where order forms on intermediate scales, but does not survive the thermodynamic limit leading to a semimetallic regime with a reduced fermionic quasiparticle weight close to the QCP. 
Lowering the temperature in the chiral Ising model leads to true condensation and the formation of long-range order, breaking the discrete $\mathbb{Z}_2$ symmetry at finite temperatures. 
This is reminiscent of the formation of magnetic domains in a ferromagnet slightly above Curie temperature, however, here in a Dirac semimetal. 
For the chiral XY and chiral Heisenberg models, no condensation occurs at any finite temperature, due to the CHMW theorem, which is fulfilled within our approach. 
This gives rise to a very large precondensation regime in the chiral XY and chiral Heisenberg case. 
The fermionic quasiparticle weight near the quantum critical point in terms of the wave function renormalizations allow us to identify non-Dirac liquid behavior, which we find to be pronounced with a strongly reduced weight close to the QCP and in the precondensation regime.
We also explored the indications for BKT physics for the chiral XY model, as far as they are accessible within our truncation.
Our unified approach for the effective chiral O($N$) models can also help to improve understanding of corresponding numerical results for microscopic models featuring such quantum critical points, e.g., spin-1/2 fermions on the honeycomb lattice near the transition to a charge-density wave or an antiferromagnetic state.
Deviations from our results in microscopic calculations, e.g., employing quantum Monte Carlo simulations, then indicate the presence of non-universal features, i.e., effects beyond true universal quantum criticality.

There are various interesting directions for future studies. 
Arguably, the most important would be to draw a clear connection to microscopic models for materials that exhibit relativistic quantum criticality, e.g., the transition in twisted WSe${}_2$ tetralayers~\cite{ma2024relativisticmotttransitionstrongly} or the one suggested for angle-tuned twisted bilayer graphene at small (but finite) temperatures.
This would imply to extend the truncation, e.g., by including different fermion and boson velocities and to define material-specific initial conditions for the flow.
In the future, it would be also interesting to explore the connection to (non-)Fermi liquids by considering a finite chemical potential, which would lead to a finite Fermi surface.
Another important aspect is to extract further thermodynamic observables from our models, e.g., transport coefficients to compare to experimental data.
This seems within reach regarding successful FRG calculations of the conductivity for relativistic $\mathrm{O}(N)$ models in $2+1$ spacetime dimensions reported in Refs.~\cite{Dupuis:2020fhh,PhysRevB.96.100501}.
It will then be very exciting to see how close the experimental data resembles the predictions from GNY models with their comparably high degree of symmetry, i.e., how well nature exhibits such universal emergent relativistic phenomena in actual solids. 

\section*{Acknowledgments}

We thank Lukas Debbeler, Nicol\`o Defenu, Kilian Fraboulet, Bilal Hawashin, Friederike Ihssen, Adrian K\"onigstein, Walter Metzner, \'Alvaro Pastor-Guti\'errez, Jan M. Pawlowski, Fabian Rennecke, and Franz R. Sattler for discussions.
M.M.S. acknowledges funding from the Deutsche Forschungsgemeinschaft (DFG, German Research Foundation) within Project-ID 277146847, SFB 1238 (project C02), and the DFG Heisenberg programme (Project-ID 452976698).
L.C. was funded by the European Union (ERC-2023-STG, Project 101115758 - QuantEmerge). Views and opinions expressed are, however, those of the author only and do not necessarily reflect those of the European Union or the European Research Council Executive Agency. Neither the European Union nor the granting authority can be held responsible for them.

\bibliography{references}

\clearpage


\onecolumngrid

\appendix

\section{Projection prescriptions for the flow equations}\label{app:Projections}

The flow equations are obtained by using the Wetterich equation \eqref{eq:WetterichEq} and the corresponding projection prescriptions
for effective action Eq.~\eqref{eq:EffAction}
and the matrix regulator $R_k$
\begin{equation}
    R_k (p,q) \equiv \begin{pmatrix}
        R_k^{(B)}(q) & 0 & 0\\
        0&0&-R_k^{(F)T}(-q)\\
        0 & R_k^{(F)}(q) &0
    \end{pmatrix}\delta_{p,q} \, .
\end{equation}

The prescription for the flow equation of the scalar potential is as follows:
\begin{equation}
    \partial_t U_k = \frac{1}{2 } \text{STr} \left[ \left(\Gamma_k^{(2)} + R_k\right)^{-1} \left(\partial_t R_k\right) \right]\bigg\rvert_{ \scriptstyle \phi_a=\mathrm{const}, \bar{\psi}=\psi=0},
    \label{eq:Potential}
\end{equation}
which leads to Eq.~\eqref{eq:PotentialFlow}. 
In addition, the flow equations  for the expansion parameters $\bar m_{\phi,k}$, $\bar \kappa_k$ and $\bar \lambda_k$ are obtained from the following projection prescriptions evaluated at the minimum of the effective potential $\rho=\rho_\mathrm{min}$
\begin{equation}
\begin{split}
    \partial_t \bar m_{\phi,k}^2 =  \partial_t u_k^\prime \big\rvert_{ \scriptstyle \rho=0} \quad \text{and} \quad \partial_t \bar \lambda_k = \partial_t u_k^{\prime\prime} \big\rvert_{ \scriptstyle \rho=0}
\end{split}
\end{equation}
for the SYM regime and
\begin{equation}
\begin{split}
     \quad \partial_t \bar \kappa_k =  -\frac{\partial_t u_k^{\prime}}{\bar\lambda_k} \bigg\rvert_{ \scriptstyle \rho=\kappa_k} \quad \text{and} \quad \partial_t \bar \lambda_k = \partial_t u_k^{\prime\prime}\big\rvert_{ \scriptstyle \rho=\kappa_k} 
\end{split}
\end{equation}
 for the SSB regime.

To obtain the projection prescription for the flow of the Yukawa coupling, we separate the bosonic field into the VEV and the fluctuations of the VEV as ${\phi_a=\phi_{a,0} + \Delta \phi_a}$. Now, the effective action $\Gamma_k^{(2)}$ can be expanded in the same fashion, with a fluctuation independent part, $\Gamma_{k,0}^{(2)} $, and a fluctuation dependent part, $\Delta \Gamma_k^{(2)}$.
In addition, as $\partial_t$ acts only on the $t$ dependence of the regulator, the Wetterich equation can be expanded by a Mercator series, 
\begin{equation}
\begin{split}
&\partial_t \Gamma_k = \frac{1}{2} \tilde{\partial}_t \mathrm{STr} \left[\ln \left(\Gamma_k^{(2)} + R_k \right)\right]\\
&\phantom{\partial_t \Gamma_k }=\frac{1}{2} \tilde{\partial}_t \mathrm{STr} \left[\ln \left(\Gamma_{k,0}^{(2)} + R_k \right)\right]+ \frac{1}{2} \tilde{\partial}_t \mathrm{STr} \sum_{n=1}^\infty \frac{(-1)^{n+1}}{n} \left[ \left(\Gamma_{k,0}^{(2)} + R_k \right)^{-1} \Delta \Gamma_k^{(2)}\right]^n,
\label{eq:MercatorSeries}
\end{split}
\end{equation}
where the operator $\tilde{\partial}_t$ acts only on the regulator dependence and reads
\begin{equation}
\begin{split}
    \tilde{\partial}_t \equiv \int_x \left[-2 x \, r_{F}^\prime(x) - \eta_{\psi} \, r_{F}(x) \right] \frac{\delta}{\delta r_{F}(x)} + \int_x \left[-2 x \, r_{B}^\prime(x) - \eta_{\phi} \, r_{B}(x) \right] \frac{\delta}{\delta r_{B}(x)}
    \end{split},
\end{equation}
where we write the regulators in the form ${R_k^{(B)}(q)=Z_{\phi,k} q^2 r_{B}(q)}$ and ${R_k^{(F)}(q)=i Z_{\psi,k} q_\mu (I_{N_\mathrm{f}}\otimes \gamma^\mu) r_{F}(q)}$.
Only the cubic-order term of Eq.~\eqref{eq:MercatorSeries} survives to the projection, which for the chiral Ising model is given by
\begin{equation}
\begin{split}
    \partial_t   h_{k,\mathrm{I}} = \frac{1}{6  N_\mathrm{f} d_\gamma} \mathrm{Tr} \left[ \frac{\overrightarrow{\delta}}{\delta \Delta \phi(q)} \frac{ \overrightarrow{\delta}}{\delta \bar\psi(q^\prime)} \tilde{\partial}_t \mathrm{STr} \left[ \left(\frac{\Delta \Gamma_k^{(2)}}{\Gamma_{k,0}^{(2)} + R_k}\right)^3\right]\frac{ \overleftarrow{\delta}}{\delta \psi(q^{\prime\prime})} \right]\Bigg\rvert_{\begin{matrix}
         \scriptstyle q=q^\prime=q^{\prime\prime}=0  \\
    \scriptstyle  \bar{\psi}=\psi=\Delta\phi=0
    \end{matrix}},
    \label{eq:PRYukawaI}
\end{split}
\end{equation}
for the chiral XY model, where the minimum lies on the $\phi_1$ axis with $\phi_{1}=\phi_{1,0}+\Delta \phi_{1}$  and  $\phi_{2}=\Delta \phi_{2}$, is 
\begin{equation}
\begin{split}
    \partial_t h_{k,\mathrm{XY}} = \frac{1}{6  N_\mathrm{f} d_\gamma} \mathrm{Tr} \left[ \left(I_2\otimes\gamma^5 \right)\frac{\overrightarrow{\delta}}{\delta \Delta \phi_2(q)} \frac{ \overrightarrow{\delta}}{\delta \bar\psi(q^\prime)} \tilde{\partial}_t \mathrm{STr} \left[ \left(\frac{\Delta \Gamma_k^{(2)}}{\Gamma_{k,0}^{(2)} + R_k}\right)^3\right]\frac{ \overleftarrow{\delta}}{\delta \psi(q^{\prime\prime})} \right]\Bigg\rvert_{\begin{matrix}
         \scriptstyle q=q^\prime=q^{\prime\prime}=0  \\
    \scriptstyle  \bar{\psi}=\psi=\Delta\phi_a=0
    \end{matrix}},
    \label{eq:PRYukawaXY}
\end{split}
\end{equation}
and for the chiral Heisenberg model, where the minimum lies on the $\phi_3$ axis with $\phi_{3}=\phi_{3,0}+\Delta \phi_{3}$  and, $\phi_{1}=\Delta \phi_{1}$ and $\phi_{2}=\Delta \phi_{2}$, is  
\begin{equation}
\begin{split}
    \partial_t  h_{k,\mathrm{H}} = \frac{1}{6  N_\mathrm{f} d_\gamma} \mathrm{Tr} \left[\left(\sigma_1\otimes I_4\right) \frac{\overrightarrow{\delta}}{\delta \Delta \phi_1(q)} \frac{ \overrightarrow{\delta}}{\delta \bar\psi(q^\prime)} \tilde{\partial}_t \mathrm{STr} \left[ \left(\frac{\Delta \Gamma_k^{(2)}}{\Gamma_{k,0}^{(2)} + R_k}\right)^3\right]\frac{ \overleftarrow{\delta}}{\delta \psi(q^{\prime\prime})} \right]\Bigg\rvert_{\begin{matrix}
         \scriptstyle q=q^\prime=q^{\prime\prime}=0  \\
    \scriptstyle  \bar{\psi}=\psi=\Delta\phi_a=0
    \end{matrix}}.
    \label{eq:PRYukawaH}
\end{split}
\end{equation}
The projection rules for the Yukawa coupling Eqs.~\eqref{eq:PRYukawaI}-\eqref{eq:PRYukawaH} evaluated at the minimum of the effective potential, i.e. $\bar\rho_{\mathrm{min},k}=0$ for the SYM regime and $\bar\rho_{\mathrm{min},k} = 
\bar\kappa_k \neq 0 $ for the SSB regime, lead to the following flow equations 
\begin{equation}
    \begin{split}
 \partial_t  \bar h^2_{k,\mathrm{I}} =& (d-4+2\eta_\psi + \eta_\phi ) \bar h_k^2 + 8 v_d \bar h_k^4    l_{11}^{(FB)d} (\tau,\omega_\psi, \omega_\phi;\eta_\psi,\eta_\phi)\\
&- 16v_d \bar\rho  \bar h_k^4  \Big[ 2  \bar  h_k^2  l_{21}^{(FB) d} (\tau,\omega_\psi , \omega_\phi;\eta_\psi,\eta_\phi) +(3 u_k'' + 2\bar\rho u_k''' )   l_{12}^{(FB) d} (\tau,\omega_\psi,\omega_\phi;\eta_\psi,\eta_\phi)\Big]
\label{eq:YukawaFlowI}
\end{split}
\end{equation}
for the chiral Ising model,
\begin{equation}
  \begin{split}
\partial_t \bar h^2_{k,\mathrm{XY}} =& (d-4+2\eta_\psi + \eta_\phi ) \bar h_k^2- 8  v_d \bar  h_k^4 \big[ l_{11}^{(FB)d} (\tau,\omega_\psi, \omega_\phi;\eta_\psi,\eta_\phi) - l_{11}^{(FB)d} (\tau,\omega_\psi, u_k' ;\eta_\psi,\eta_\phi) \big] \\
&- 32 v_d \bar\rho \bar  h_k^4    u_k'' l_{111}^{(FBB) d} (\tau,\omega_\psi, u_k', \omega_\phi;\eta_\psi,\eta_\phi) 
\label{eq:YukawaFlowXY}
\end{split}
\end{equation}
for the chiral XY model and
\begin{equation}
    \begin{split}
\partial_t \bar h^2_{k,\mathrm{H}} =& (d-4+2\eta_\psi + \eta_\phi ) \bar h_k^2 - 8  v_d \bar  h_k^4  l_{11}^{(FB)d} (\tau,\omega_\psi, \omega_\phi;\eta_\psi,\eta_\phi) - 32 v_d \bar\rho \bar  h_k^4    u_k'' l_{111}^{(FBB) d} (\tau,\omega_\psi, u_k', \omega_\phi;\eta_\psi,\eta_\phi) 
\label{eq:YukawaFlowH}
\end{split}
\end{equation}
for the chiral Heisenberg model.  
The threshold functions $l_{n_1 n_2}^{(FB)d}$ and $l_{n_1 n_2 n_3}^{(FBB)d}$ are defined below and depend on the regulator choice.

Finally, we need the projection rules for the wave function renormalizations to obtain the expressions of the anomalous dimensions. To that end, we use again the Taylor expansion for the Wetterich equation given in Eq.~\eqref{eq:MercatorSeries}. 
After the projection, only the second-order term survives. 
For the chiral Ising model, we project on the radial mode
\begin{equation}
\begin{split}
    \partial_t Z_{\phi,\mathrm{I}} = -\frac{1}{4}  \frac{\partial}{\partial q^2} \int \frac{d^d q^\prime}{(2\pi)^d} \frac{\overrightarrow{\delta}}{\delta \Delta\phi(-q)} \tilde{\partial}_t \mathrm{STr} \left[ \left(\frac{\Delta \Gamma_k^{(2)}}{\Gamma_{k,0}^{(2)} + R_k}\right)^2\right] \frac{\overleftarrow{\delta}}{\delta \Delta\phi(q^\prime)} \Bigg\rvert_{\begin{matrix}
         \scriptstyle q=q^\prime=0  \\
    \scriptstyle  \bar{\psi}=\psi=\Delta\phi=0
    \end{matrix}},
    \label{eq:ADI}
\end{split}
\end{equation} 
while for the chiral XY model
\begin{equation}
\begin{split}
    \partial_t Z_{\phi,\mathrm{XY}} = -\frac{1}{4}  \frac{\partial}{\partial q^2} \int \frac{d^d q^\prime}{(2\pi)^d} \frac{\overrightarrow{\delta}}{\delta \Delta\phi_2(-q)} \tilde{\partial}_t \mathrm{STr} \left[ \left(\frac{\Delta \Gamma_k^{(2)}}{\Gamma_{k,0}^{(2)} + R_k}\right)^2\right] \frac{\overleftarrow{\delta}}{\delta \Delta\phi_2(q^\prime)} \Bigg\rvert_{\begin{matrix}
         \scriptstyle q=q^\prime=0  \\
    \scriptstyle  \bar{\psi}=\psi=\Delta\phi_a=0
    \end{matrix}},
    \label{eq:ADXY}
\end{split}
\end{equation} 
and for the chiral Heisenberg model
\begin{equation}
\begin{split}
    \partial_t Z_{\phi,\mathrm{H}} = -\frac{1}{4}  \frac{\partial}{\partial q^2} \int \frac{d^d q^\prime}{(2\pi)^d} \frac{\overrightarrow{\delta}}{\delta \Delta\phi_{1}(q)} \tilde{\partial}_t \mathrm{STr} \left[ \left(\frac{\Delta \Gamma_k^{(2)}}{\Gamma_{k,0}^{(2)} + R_k}\right)^2\right] \frac{\overleftarrow{\delta}}{\delta \Delta\phi_{1}(q^\prime)} \Bigg\rvert_{\begin{matrix}
         \scriptstyle q=q^\prime=0  \\
    \scriptstyle  \bar{\psi}=\psi=\Delta\phi_a=0
    \end{matrix}},
    \label{eq:ADH}
\end{split}
\end{equation}
we project on the Goldstone mode.
Note that the expressions are different if we project on the radial mode or on the Goldstone mode(s).
In addition, the projection prescription for the fermionic wave function renormalization
   \begin{equation}
\begin{split}
    \partial_t Z_\psi = \frac{i}{4 N_\mathrm{f} d d_\gamma} \mathrm{Tr}\left[(I_2\otimes\gamma^\mu) \frac{\partial}{\partial q^\mu} \int \frac{d^d q^\prime}{(2\pi)^d} \frac{\overrightarrow{\delta}}{\delta \bar{\psi}(-q)} \tilde{\partial}_t \mathrm{STr} \left[ \left(\frac{\Delta \Gamma_k^{(2)}}{\Gamma_{k,0}^{(2)} + R_k}\right)^2\right]\frac{\overleftarrow{\delta}}{\delta \psi(q^\prime)} \right]\Bigg\rvert_{\begin{matrix}
         \scriptstyle q=q^\prime=0  \\
    \scriptstyle  \bar{\psi}=\psi=\Delta\phi_a=0
    \end{matrix}}.
    \label{eq:ADF}
\end{split}
\end{equation}
For $N\in\{1,2,3\}$, the 
implicit equation for the boson anomalous dimension $\eta_\phi$ is given by
\begin{equation}
\begin{split}
\eta_\phi =& 8 \frac{v_d}{d} N_\mathrm{f} d_\gamma \bar h_k^2 \Big[ m_{4}^{(F)d} (\tau, \omega_\psi;\eta_\psi) - 2(1-2\delta_{3,N}) \bar \rho \bar h_k^2 m_{2}^{(F)d} (\tau, \omega_\psi;\eta_\psi)\Big]\\
&+ \delta_{1,N} 8 \frac{v_d}{d} \bar \rho (3u_k''+2\bar\rho u_k''')^2 m_{40}^{(B) d}(\tau,\omega_\phi,0;\eta_\phi)+ (\delta_{2,N}+\delta_{3,N})16 \frac{v_d}{d} \bar \rho (u_k'')^2 m_{22}^{(B) d}(\tau,u_k',\omega_\phi;\eta_\phi)\,,
\label{eq:AnomalousDimPhiFlow}
\end{split}
\end{equation}
projected on the radial mode for the chiral Ising and on the Goldstone mode otherwise, and the one for the fermion anomalous dimension $\eta_\psi$ reads
\begin{equation}
\begin{split}
\eta_\psi =&\ 8 \frac{v_d}{d} \bar h^2_k m_{12}^{(FB)d} (\tau,\omega_\psi, \omega_\phi;\eta_\psi,\eta_\phi) + 8 (N-1)   \frac{v_d}{d} \bar h^2_k m_{12}^{(FB)d} (\tau,\omega_\psi , u';\eta_\psi,\eta_\phi)\,.
\label{eq:AnomalousDimPsiFlow}
\end{split}
\end{equation}
The threshold functions $m_2^{(F)d}$, $m_4^{(F)d}$, $m_{n_1n_2}^{(B)d}$ and $m_{12}^{(FB)d}$ are given below.

\section{Threshold functions}\label{app:ThresholdFunctions}

The threshold functions, which involve the bosonic and fermionic loops and the regulator dependence, are given in our flow equations at finite temperature by
\begin{equation}
\begin{split}
    &l_n^{(B)d}(\tau,\omega;\eta_\phi) = \frac{\delta_{n,0}+n}{2} \frac{v_{d-1}}{v_d} \frac{\tau}{2\pi} \sum_{n\in \mathbb{Z}} \int_0^\infty \mathrm{d}y \, y^{\frac{d-3}{2}} y_B \left(P_B (y_B) + \omega \right)^{-(n+1)} \frac{\partial_t \left(Z_{\phi,k} r_{B}(y_B) \right) }{Z_{\phi,k}}, \\
   &l_n^{(F)d}(\tau,\omega;\eta_\psi) = \left(\delta_{n,0}+n\right) \frac{v_{d-1}}{v_d} \frac{\tau}{2\pi} \sum_{n\in \mathbb{Z}} \int_0^\infty \mathrm{d}y \, y^{\frac{d-3}{2}} y_F \left(P_F (y_F) + \omega \right)^{-(n+1)} (1+r_{F}(y_F)) \frac{\partial_t \left(Z_{\psi,k} r_{F}(y_F) \right) }{Z_{\psi,k}},\\
    &l_{n_1 n_2}^{(FB)d}(\tau,\omega_1,\omega_2;\eta_\psi,\eta_\phi) = -\frac{1}{2} \frac{v_{d-1}}{v_d} \frac{\tau}{2\pi} \sum_{n \in \mathbb{Z}} \int_0^\infty \mathrm{d}y \, y^{\frac{d-3}{2}} \tilde{\partial}_t \left[ \frac{1}{\left(P_F (y_F) + \omega_1 \right)^{n_1} \left(P_B (y_B) + \omega_2 \right)^{n_2}} \right],\\
    &l_{n_1 n_2 n_3}^{(FBB)d}(\tau,\omega_1,\omega_2,\omega_3;\eta_\psi,\eta_\phi) = -\frac{1}{2} \frac{v_{d-1}}{v_d} \frac{\tau}{2\pi} \sum_{n \in \mathbb{Z}} \int_0^\infty \mathrm{d}y \, y^{\frac{d-3}{2}} \tilde{\partial}_t \left[ \frac{1}{\left(P_F (y_F) + \omega_1 \right)^{n_1} \left(P_B (y_B) + \omega_2 \right)^{n_2} \left(P_B (y_B) + \omega_3 \right)^{n_3} } \right],\\
    &m_2^{(F)d}(\tau,\omega;\eta_\psi) = -\frac{1}{2} \frac{v_{d-1}}{v_d} \frac{d}{d-1} \frac{\tau}{2\pi} \sum_{n \in \mathbb{Z}} \int_0^\infty \mathrm{d}y \, y^{\frac{d-1}{2}} \tilde{\partial}_t \left[\frac{\dot{P}_F (y_F)}{\left(P_F (y_F) + \omega \right)^{2} } \right]^{2}, \\
    &m_4^{(F)d}(\tau,\omega;\eta_\psi) = -\frac{1}{2} \frac{v_{d-1}}{v_d} \frac{d}{d-1} \frac{\tau}{2\pi} \sum_{n \in \mathbb{Z}} \int_0^\infty \mathrm{d}y \, y^{\frac{d-1}{2}} y_F \, \tilde{\partial}_t \left[ \frac{\partial}{\partial y} \left(\frac{1 + r_F (y_F)}{P_F (y_F) + \omega }\right) \right]^{2},  \\
    &m_{n_1 n_2}^{(B)d}(\tau,\omega_1,\omega_2;\eta_\phi) = -\frac{1}{2} \frac{v_{d-1}}{v_d} \frac{d}{d-1} \frac{\tau}{2\pi} \sum_{n \in \mathbb{Z}} \int_0^\infty \mathrm{d}y \, y^{\frac{d-1}{2}} \tilde{\partial}_t \left[\frac{\dot{P}_B (y_B)}{\left(P_B (y_B) + \omega_1 \right)^{n_1} } \frac{\dot{P}_B (y_B)}{\left(P_B (y_B) + \omega_2 \right)^{n_2} }  \right],\\
   &m_{n_1 n_2}^{(FB)d} (\tau, \omega_1, \omega_2; \eta_\psi, \eta_\phi) = -\frac{1}{2} \frac{v_{d-1}}{v_d} \frac{d}{d-1} \frac{\tau}{2\pi} \sum_{n \in \mathbb{Z}} \int_0^\infty \mathrm{d}y \, y^{\frac{d-1}{2}} \tilde{\partial}_t \left[ \frac{1 + r_F (y_F)}{\left(P_F (y_F) + \omega_1\right)^{n_1} }  \frac{\dot{P}_B (y_B)}{\left(P_B (y_B) + \omega_2 \right)^{n_2} } \right],
\end{split}
\end{equation} 
where we introduced $y\equiv q^2/k^2$ and the dot notation denotes derivative with respect to $y$. 
The Matsubara frequencies and the dimensionless temperature parameter, $\tau$, are included in $y_{B/F}=y+ \left(\tau c_{B/F,n}\right)^2$ with ${c_{B,n}=n}$ and ${c_{F,n}=(n+1/2)}$.
We define the bosonic and fermionic functions $P_B(x)=x(1+r_{B}(x))$ and $P_F(x)=x(1+r_{F}(x))^2$.

Choosing the Litim regulator,  
   $ {r_{B}(x) = \left(x^{-1} - 1\right) \theta(1-x) }$
and
${r_{F}(x) = \left(x^{-1/2} - 1\right) \theta(1-x)}$,
the thermal threshold functions from above read
\begin{equation}
\begin{split}
    &l_n^{(B)d}(\tau,\omega;\eta_\phi) = \left(s_0^d(\tau) - \eta_\phi \hat{s}_0^d(\tau)\right) l_n^{(B)d}(\omega;0),\\
    &l_n^{(F)d}(\tau,\omega;\eta_\psi) = \left(s_0^{(F)d}(\tau) - \eta_\psi \hat{s}_0^{(F)d}(\tau)\right) l_n^{(F)d}(\omega;0),\\
   & l_{n_1 n_2}^{(FB)d}(\tau,\omega_1,\omega_2;\eta_\psi,\eta_\phi) = \frac{1}{\left(1+\omega_1\right)^{n_1}}  \frac{1}{\left(1+\omega_2\right)^{n_2}}\\
    &\phantom{l_{n_1 n_2}^{(FB)d} }\times
     \Big\{ n_1\left[ s_0^{(F)d}(\tau) -\eta_\psi \hat{s}_0^{(F)d}(\tau) \right] l_0^{(F) d}(\omega_1;0) + n_2 \left[ s^{(F)d}_0(\tau) - \eta_\phi \hat{s}^{d}(\tau) \right]  l_0^{(B) d}(\omega_2;0) \Big\}, \\
    &l_{n_1 n_2 n_3}^{(FBB)d}(\tau,\omega_1,\omega_2,\omega_3;\eta_\psi,\eta_\phi) =   \frac{1}{\left(1+\omega_1\right)^{n_1}}  \frac{1}{\left(1+\omega_2\right)^{n_2}} \frac{1}{(1+\omega_3)^{n_3}}\\
    &\phantom{l_{n_1 n_2 n_3}^{(FBB)d}} \times\Big\{ n_1\left[ s_0^{(F)d}(\tau) -\eta_\psi \hat{s}_0^{(F)d}(\tau) \right] l_0^{(F) d}(\omega_1;0) + \left[ s^{(F)d}_0(\tau) - \eta_\phi \hat{s}^{d}(\tau) \right] \left[n_2  l_0^{(B) d}(\omega_2;0) +n_3  l_0^{(B) d}(\omega_3;0)  \right] \Big\},\\
    &m_2^{(F)d}(\tau,\omega;\eta_\psi) = s_0^{(F)d} (\tau) m_2^{(F)d}(\omega;0),\\
    &m_4^{(F)d}(\tau,\omega;\eta_\psi) = s_0^{(F)d} (\tau) \frac{1}{\left(1+\omega\right)^4} + \frac{1-\eta_\psi}{2} t_4 (\tau) \frac{1}{\left(1+\omega\right)^3} -\left[\frac{1-\eta_\psi}{4} t_4 (\tau) + \frac{1}{4} s_0^{(F)d} (\tau) \right]\frac{1}{\left(1+\omega\right)^2} ,\\
    &m_{n_1 n_2}^{(B)d}(\tau,\omega_1,\omega_2;\eta_\phi) = s_0^{d} (\tau) m_{n_1 n_2}^{(B)d}(\omega_1,\omega_2;0),\\
    &m_{12}^{(FB)d}(\tau,\omega_1,\omega_2;\eta_\psi,\eta_\phi) = \Big[ s_0^{(F)d}(\tau) - \eta_\phi \hat{t}^{(FB)d} (\tau) \Big] m_{12}^{(FB) d} (\omega_1, \omega_2;0,0).
    \label{eq:ThresholdFunctionsT}
\end{split}
\end{equation}

The threshold functions at zero temperature are given by
\begin{equation}
\begin{split}
    &l_n^{(B)d}(\omega;\eta_\phi) = \left(\delta_{n,0}+n\right) \frac{2}{d} \left(1-\frac{\eta_\phi}{d+2}\right) \frac{1}{\left(1+\omega\right)^{n+1}}, \\
    &l_n^{(F)d}(\omega;\eta_\psi) = \left(\delta_{n,0}+n\right) \frac{2}{d} \left(1-\frac{\eta_\psi}{d+1}\right) \frac{1}{\left(1+\omega\right)^{n+1}},\\
    &l_{n_1 n_2}^{(FB)d}(\omega_1,\omega_2;\eta_\psi,\eta_\phi) = \frac{2}{d} \frac{1}{\left(1+\omega_1\right)^{n_1}}\frac{1}{\left(1+\omega_2\right)^{n_2}} \left[\frac{n_1}{1+\omega_1} \left(1-\frac{\eta_\psi}{d+1}\right)  + \frac{n_2}{1+\omega_2} \left(1-\frac{\eta_\phi}{d+2}\right) \right], \\
    &l_{n_1 n_2 n_3}^{(FBB)d}(\omega_1,\omega_2,\omega_3;\eta_\psi,\eta_\phi) = \frac{2}{d} \frac{1}{\left(1+\omega_1\right)^{n_1}}\frac{1}{\left(1+\omega_2\right)^{n_2}} \frac{1}{\left(1+\omega_3\right)^{n_3}} \\
    &\phantom{l_{n_1 n_2 n_3}^{(FBB)d}(\omega_\psi,\omega_\varphi,\omega_\theta;\eta_\psi,\eta_\phi) = }\times\left[\frac{n_1}{1+\omega_1} \left(1-\frac{\eta_\psi}{d+1}\right)  + \left(\frac{n_2}{1+\omega_2} +\frac{n_3}{1+\omega_3}\right) \left(1-\frac{\eta_\phi}{d+2}\right) \right], \\
    &m_2^{(F)d}(\omega;\eta_\psi) = \frac{1}{\left(1+\omega\right)^{4}}, \\
   & m_4^{(F)d}(\omega;\eta_\psi) = \frac{1}{\left(1+\omega\right)^{4}} + \frac{1-\eta_\psi}{d-2} \frac{1}{\left(1+\omega\right)^{3}} - \left(\frac{1-\eta_\psi}{2d-4} + \frac{1}{4}\right) \frac{1}{\left(1+\omega\right)^{2}} , \\
   & m_{n_1 n_2}^{(B)d}(\omega_1,\omega_2;\eta_\phi) = \frac{1}{\left(1+ \omega_1\right)^{n_1}} \frac{1}{\left(1+ \omega_2\right)^{n_2}} , \\
   & m_{n_1 n_2}^{(FB)d}(\omega_1,\omega_2;\eta_\psi,\eta_\phi) = \left(1-\frac{\eta_\phi}{d+1}\right)\frac{1}{\left(1+ \omega_1\right)^{n_1}} \frac{1}{\left(1+ \omega_2\right)^{n_2}},
\end{split}
\end{equation}
and the thermal Matsubara sums from Eq.~\eqref{eq:ThresholdFunctionsT} are
\begin{equation}
\begin{split}
    &s_0^{d} (\tau)= \frac{v_{d-1}}{v_d} \frac{d}{d-1} \frac{\tau}{2\pi} \sum_{n\in \mathbb{Z}} \theta\left(1-\left(\tau c_{B,n}\right)^2\right) \left(1-\left(\tau c_{B,n}\right)^2\right)^{\frac{d-1}{2}},\\
    &\hat{s}_0^{d} (\tau)= \frac{v_{d-1}}{v_d} \frac{d}{d^2-1} \frac{\tau}{2\pi} \sum_{n\in \mathbb{Z}} \theta\left(1-\left(\tau c_{B,n}\right)^2\right) \left(1-\left(\tau c_{B,n}\right)^2\right)^{\frac{d+1}{2}},\\
    &s_0^{(F)d} (\tau)=  \frac{v_{d-1}}{v_d} \frac{d}{d-1} \frac{\tau}{2\pi} \sum_{n\in \mathbb{Z}} \theta\left(1-\left(\tau c_{F,n}\right)^2\right) \left(1-\left(\tau c_{F,n}\right)^2\right)^{\frac{d-1}{2}},\\
    &\hat{s}_0^{(F)d} (\tau)= \frac{v_{d-1}}{v_d} \frac{d}{d-1} \frac{\tau}{2\pi} \sum_{n\in \mathbb{Z}} \theta\left(1-\left(\tau c_{F,n}\right)^2\right) \left(1-\left(\tau c_{F,n}\right)^2\right)^{\frac{d-1}{2}}\\
     &\phantom{\hat{s}_0^{(F)d} (\tau)= \frac{v_{d-1}}{v_d} \frac{d}{d-1} \frac{\tau}{2\pi} \sum_{n\in \mathbb{Z}}}\times\left[1- \abs{\tau c_{F,n}}^{-1} {}_2F_1\left(-\frac{1}{2},\frac{d-1}{2};\frac{d+1}{2}; \frac{\left(\tau c_{F,n}\right)^2-1}{\left(\tau c_{F,n}\right)^2}\right)\right],\\
     &\hat{s}^{d} (\tau)= \frac{v_{d-1}}{v_d} \frac{d}{d^2 - 1} \frac{\tau}{2\pi} \sum_{n\in \mathbb{Z}} \theta\left(1-\left(\tau c_{F,n}\right)^2\right) \left(1-\left(\tau c_{F,n}\right)^2\right)^{\frac{d+1}{2}},\\
     &\hat{t}^{(FB)d} (\tau)= \frac{v_{d-1}}{v_d} \frac{d}{d^2-1} \frac{\tau}{2\pi} \sum_{n\in \mathbb{Z}} \theta\left(1-\left(\tau c_{F,n}\right)^2\right) \left(1-\left(\tau c_{F,n}\right)^2\right)^{\frac{d+1}{2}} \abs{\tau c_{F,n}}^{-1} {}_2 F_1\left(\frac{1}{2}, \frac{d+1}{2}; \frac{d+3}{2}; \frac{(\tau c_{F,n})^2 -1}{(\tau c_{F,n})^2}\right),\\ 
     &t_4 (\tau)= \frac{v_{d-1}}{v_d} \frac{d}{d-1} \frac{\tau}{2\pi} \sum_{n\in \mathbb{Z}} \theta\left(1-\left(\tau c_{F,n}\right)^2\right) \left(1-\left(\tau c_{F,n}\right)^2\right)^{\frac{d+1}{2}} \abs{\tau c_{F,n}}^{-2} \\
     &\phantom{t_4 (\tau)=\frac{v_{d-1}}{v_d} \frac{d}{d-1} \frac{\tau}{2\pi} \sum_{n\in \mathbb{Z}}}\times\left[1 - \abs{\tau c_{F,n}}^{-2} \frac{d-1}{d+1} {}_2 F_1\left(\frac{1}{2}, \frac{d+1}{2}; \frac{d+3}{2}; \frac{(\tau c_{F,n})^2 -1}{(\tau c_{F,n})^2}\right)\right],
\end{split}
\end{equation}    
whose closed expressions in $d=2+1$ dimensions are given by
\begin{equation}
\begin{split}
    &s_0^{d} (\tau)= \frac{v_{d-1}}{v_d} \frac{d}{d-1} \frac{\tau}{2\pi} \frac{1}{3} \left(1+ 2 s_B(\tau)\right)\left(3-\tau^2 s_B(\tau) - \tau^2 s_B(\tau)^2\right),\\
    &\hat{s}_0^{d} (\tau)= \frac{v_{d-1}}{v_d} \frac{d}{d^2-1} \frac{\tau}{2\pi} \frac{1}{15} \left(1+ 2 s_B(\tau)\right) \left[15+ \tau^2 s_B(\tau) \left(1+s_B(\tau)\right) \left(-10 - \tau^2 + 3\tau^2 s_B(\tau) + 3\tau^2 s_B(\tau)^2 \right) \right],\\
    &s_0^{(F)d} = \frac{v_{d-1}}{v_d} \frac{d}{d-1} \frac{\tau}{2\pi} \frac{1}{6} s_F(\tau) \left(12 + \tau^2 - 4 \tau^2 s_F(\tau)^2\right) ,\\
    &\hat{s}_0^{(F)d} (\tau)= \frac{v_{d-1}}{v_d} \frac{d}{d-1} \frac{\tau}{2\pi} \frac{1}{6}  s_F(\tau) \left[ 4 + \tau^2 + \tau^2 s_F(\tau) \left(-\tau -4s_F(\tau)+ 2 \tau s_F(\tau)^2\right)\right],\\
     &\hat{s}^{d} (\tau)= \frac{v_{d-1}}{v_d} \frac{d}{d^2 - 1} \frac{\tau}{2\pi} \frac{1}{120} s_F(\tau) \left[240 + 40 \tau^2 + 7\tau^4 - 40 \tau^2 s_F(\tau)^2 (4 + \tau^2) + 48 \tau^4 s_F(\tau)^4\right],\\
     &\hat{t}^{(FB)d} (\tau)= \frac{v_{d-1}}{v_d} \frac{d}{d^2-1} \frac{\tau}{2\pi} \frac{2}{3} s_F(\tau) \left[4  + \tau^2 + \tau^2 s_F(\tau)\left(-\tau - 4s_F(\tau) + 2 \tau s_F(\tau)^2\right)\right],\\ 
     &t_4 (\tau)= \frac{v_{d-1}}{v_d} \frac{d}{d-1} \frac{\tau}{2\pi} \Bigg\{\frac{1}{6} s_F(\tau) \left(-12 -\tau^2 + 4\tau^2 s_F(\tau)^2\right) + \frac{1}{3} \Bigg[32 s_F(\tau) + 2\tau s_F(\tau)^2 (-12 -\tau^2) + 4\tau^3 s_F(\tau)^4 \\
     &\phantom{t_4 (\tau)=}- \frac{1}{20 } \tau \left(-32 s_F(\tau) + 2\tau \left(40 + 20 \tau^2 + 7 \tau^4\right) s_F(\tau)^2 + 40 \tau^3(-2 - \tau^2) s_F(\tau)^4 + 32 \tau^5 s_F(\tau)^6\right)\Bigg] + ...\Bigg\},\
\end{split}
\end{equation}
where $s_F(\tau)=\lfloor \frac{1}{\tau} + \frac{1}{2}\rfloor$ and $s_B(\tau)=\lfloor \frac{1}{\tau}\rfloor$.
Here, we used the Poisson resummation formula
\begin{equation}
    \sum_{n\in \mathbb{Z}} f(n) = \sum_{l\in \mathbb{Z}} \int_\mathbb{R} \mathrm{d} q f(q) e^{-2\pi i q l}
\end{equation}
and the identity
\begin{equation}
    \mathrm{Li}_n (-z) - (-1)^{n-1}  \mathrm{Li}_n (-1/z) = -\frac{(2\pi i )^n}{n!} B_n\left(\frac{\ln z}{2\pi i} + \frac{1}{2}\right),
\end{equation}
for $z\not\in (0,1)$, where $B_n(z)$ are the Bernoulli polynomials and $ \mathrm{Li}_n (z)$ denotes the $n$th polylogarithm. With
\begin{equation}
    \begin{split}
        \ln\left(e^{-2\pi i /\tau}\right) &= -2\pi i + 2\pi i s_F (\tau),\\
        \ln\left(-e^{-2\pi i /\tau}\right) &=
        \pi i -2\pi i + 2\pi i s_B (\tau).
    \end{split}
\end{equation}
The Matsubara sum $t_4$ is only an approximation, as it was not possible to obtain the analytical result.

Note that for the mixed threshold functions $l^{(FB)d}_{n_1 n_2}$, $l^{(FB)d}_{n_1 n_2 n_3}$ and $m^{(FB)d}_{1 2}$, we used an approximation following the prescription outlined in Ref.~\cite{Braun_2012}.
Specifically, the Matsubara frequencies were modified via the replacement $(\tau c_{B,n})^2 \rightarrow (\tau c_{F,n})^2$, which allows for a simplification and a good approximation of the threshold functions.
The exact results for the mixed threshold functions are given here
\begin{equation}
\begin{split}
    l_{n_1 n_2}^{(FB)d}&(\tau,\omega_1,\omega_2;\eta_\psi,\eta_\phi) =-\frac{1}{2} \frac{v_{d-1}}{v_d} \frac{d}{d-1} \frac{\tau}{2\pi} \frac{1}{(1+\omega_1)^{n_1} (1+\omega_2)^{n_2}} \\
    &\times\Bigg\{ \sum_{n=0}^{\lfloor\frac{1}{\tau}+\frac{1}{2}\rfloor-1} \int_0^{1-(\tau c_{F,n})^2} \mathrm{d}y \,y^{\frac{d-3}{2}} \left[ \frac{-2n_1}{1+\omega_1} \left[1+\eta_\psi (y_F^{1/2} -1 )\right] + \frac{-n_2}{1+\omega_2} \left[2+\eta_\phi (y_B -1 )\right] \right] \\
   &+\sum_{n=-\lfloor\frac{1}{\tau}\rfloor}^{-1} \int_0^{1-(\tau c_{B,n})^2} \mathrm{d}y \,y^{\frac{d-3}{2}} \left[ \frac{-2n_1}{1+\omega_1} \left[1+\eta_\psi (y_F^{1/2} -1 )\right] + \frac{-n_2}{1+\omega_2} \left[2+\eta_\phi (y_B -1 )\right] \right]\\
   &+ \sum_{n=0}^{\lfloor\frac{1}{\tau}+\frac{1}{2}\rfloor-1} \int_{1-(\tau c_{F,n})^2}^{1-(\tau c_{B,n})^2} \mathrm{d}y \,y^{\frac{d-3}{2}} \frac{-n_2}{y_B+\omega_2} \left[2+\eta_\phi (y_B -1 )\right] \\
    &+\sum_{n=-\lfloor\frac{1}{\tau}\rfloor}^{-1} \int_{1-(\tau c_{B,n})^2}^{1-(\tau c_{F,n})^2} \mathrm{d}y \, y^{\frac{d-3}{2}} \frac{-2n_1}{y_F+\omega_1} \left[1+\eta_\psi (y_F^{1/2} -1 )\right]\Bigg\},
    \label{eq:ln1n2Exact}
\end{split}
\end{equation}
\begin{equation}
\begin{split}
    l_{n_1 n_2 n_3}^{(FBB)d}&(\tau,\omega_1,\omega_2,\omega_3;\eta_\psi,\eta_\phi) =-\frac{1}{2} \frac{v_{d-1}}{v_d} \frac{d}{d-1} \frac{\tau}{2\pi} \frac{1}{(1+\omega_1)^{n_1} (1+\omega_2)^{n_2} (1+\omega_3)^{n_3}} \\
    &\times\Bigg\{ \sum_{n=0}^{\lfloor\frac{1}{\tau}+\frac{1}{2}\rfloor-1} \int_0^{1-(\tau c_{F,n})^2} \mathrm{d}y \,y^{\frac{d-3}{2}} \left[ \frac{-2n_1}{1+\omega_1} \left[1+\eta_\psi (y_F^{1/2} -1 )\right] + \left( \frac{-n_2}{1+\omega_2} + \frac{-n_3}{1+\omega_3}\right) \left[2+\eta_\phi (y_B -1 )\right] \right] \\
   &+\sum_{n=-\lfloor\frac{1}{\tau}\rfloor}^{-1} \int_0^{1-(\tau c_{B,n})^2} \mathrm{d}y \,y^{\frac{d-3}{2}} \left[ \frac{-2n_1}{1+\omega_1} \left[1+\eta_\psi (y_F^{1/2} -1 )\right] + \left( \frac{-n_2}{1+\omega_2} + \frac{-n_3}{1+\omega_3}\right) \left[2+\eta_\phi (y_B -1 )\right] \right]\\
   &+ \sum_{n=0}^{\lfloor\frac{1}{\tau}+\frac{1}{2}\rfloor-1} \int_{1-(\tau c_{F,n})^2}^{1-(\tau c_{B,n})^2} \mathrm{d}y \,y^{\frac{d-3}{2}} \left( \frac{-n_2}{y_B +\omega_2} + \frac{-n_3}{y_B +\omega_3}\right)  \left[2+\eta_\phi (y_B -1 )\right] \\
    &+\sum_{n=-\lfloor\frac{1}{\tau}\rfloor}^{-1} \int_{1-(\tau c_{B,n})^2}^{1-(\tau c_{F,n})^2} \mathrm{d}y \, y^{\frac{d-3}{2}} \frac{-2n_1}{y_F+\omega_1} \left[1+\eta_\psi (y_F^{1/2} -1 )\right]\Bigg\},
    \label{eq:ln1n2n3Exact}
\end{split}
\end{equation}
and
\begin{equation}
\begin{split}
    m_{12}^{(FB)d}&(\tau,\omega_1,\omega_2;\eta_\psi,\eta_\phi) =-\frac{1}{2} \frac{v_{d-1}}{v_d} \frac{d}{d-1} \frac{\tau}{2\pi} \\
    &\times\Bigg\{
    \sum_{n=-\lfloor\frac{1}{\tau}\rfloor}^{-1} \int_{1-(\tau c_{B,n})^2}^{1-(\tau c_{F,n})^2} \mathrm{d}y \, \frac{y^{\frac{d-1}{2}}}{(1+\omega_1)(y_B + \omega_2)^2} \left[y_F^{-1/2} - \eta_\psi (y_F^{-1/2} - 1) \right] \left(1-\frac{2}{1+\omega_1}\right)\\
   &+\eta_\phi \Bigg[ \sum_{n=0}^{\lfloor\frac{1}{\tau}+\frac{1}{2}\rfloor-1} \int_0^{1-(\tau c_{F,n})^2} \mathrm{d}y \, \frac{y^{\frac{d-1}{2}}}{(1+\omega_1)(1+\omega_2)^2} y_F^{-1/2} + \sum_{n=-\lfloor\frac{1}{\tau}\rfloor}^{-1} \int_0^{1-(\tau c_{B,n})^2} \mathrm{d}y \, \frac{y^{\frac{d-1}{2}}}{(1+\omega_1)(1+\omega_2)^2} y_F^{-1/2}\\
   &+ \sum_{n=0}^{\lfloor\frac{1}{\tau}+\frac{1}{2}\rfloor-1} \int_{1-(\tau c_{F,n})^2}^{1-(\tau c_{B,n})^2} \mathrm{d}y \, \frac{y^{\frac{d-1}{2}}}{(y_F+\omega_1)(1+\omega_2)^2} \Bigg] \\
    & - 2 \sum_{n=-\lfloor\frac{1}{\tau}\rfloor}^{-1}  \frac{(1-(\tau c_{B,n})^2)^{\frac{d-1}{2}} \left(1-(\tau c_{B,n})^2 + (\tau c_{F,n})^2\right)^{-1/2}}{(1+\omega_1)(1+\omega_2)^2}\\
    &- 2 \sum_{n=0}^{\lfloor\frac{1}{\tau}+\frac{1}{2}\rfloor-1} \frac{(1-(\tau c_{B,n})^2)^{\frac{d-1}{2}}}{(1+(\tau c_{F,n})^2 - (\tau c_{B,n})^2  +\omega_1)(1+\omega_2)^2}\Bigg\}.
    \label{eq:m12Exact}
\end{split}
\end{equation}



\section{Phase diagram for different ultraviolet Yukawa couplings}\label{app:YukawaCouplings}

We have also performed calculations with different initial conditions for $h^2$ and found
that the position of the quantum critical point is shifted, but the universal quantities and scaling exponents remain unaffected. The qualitative behavior of the phase diagram remains the same.
Different choices of the initial conditions for the effective model correspond to different microscopic model.
The critical point slightly moves along the coupling axis and the non-universal $A_\pm$ and their ratios $A_+/A_-$ also shift. 
For example, in the chiral Ising model with an initial $\bar h^2=2$, the non-universal prefactors are $A_+ = 0.173$ and $A_- = 0.132$, with $A_+ / A_- = 1.300$, and the critical coupling $g_c$ is at $5.035$, while with an initial $\bar h^2=5$, we find $A_+ = 0.153$ and $A_- = 0.097$, with $A_+ / A_- = 1.578$, and the $g_c$ is at $5.238$.

\section{Scaling region \texorpdfstring{$a=2$}{a=2}}\label{app:a2}

To understand the existence of the scaling region at very low temperatures $T$ and couplings smaller than the critical one, $g<g_c$, we study the dimensionful flow equation for the bosonic mass in the SYM phase at the infrared regime, where the anomalous dimensions vanish, $\eta_\phi\rvert_{k\rightarrow0}=\eta_\psi\rvert_{k\rightarrow0}=0$,
\begin{equation}
    \begin{split}
    \partial_k m^2_{\phi,k} \Big\rvert_{k\rightarrow0} \simeq& \frac{8}{3\pi^2} h_{k}^2 s_0^{(F) d}(\tau) - \frac{N+2}{6\pi^2} \frac{\lambda_{k}}{\left(1+ k^{-2 }m^2_{\phi,{k}}\right)^2}  s_0^{d}(\tau)\Bigg\rvert_{k\rightarrow0}.
    \label{eq:DimensionfulLambda1}
    \end{split}
\end{equation}
At the infrared limit ($k\rightarrow0$) and very low temperatures, the reduced temperature $\tau=2\pi T/k$ can remain finite. For $\tau<2$ the fermionic Matsubara sum contributes, which can be approximated by $\theta(k-2\pi T)$.
In addition, at the infrared regime, the prefactor of the bosonic Matsubara sum ${\lambda_{k}/\left(1+ k^{-2 }m^2_{\phi,k}\right)^2\sim k^4}$ vanishes.
As a result, the flow equation reduces to $\partial_k m^2_{\phi,k} \simeq \frac{8}{3\pi^2} h^2 \theta(k-2\pi T) + \mathcal{O}(k^4)$.
Integrating this equation leads to
\begin{equation}
    \begin{split}
   1/\xi_T = m_{\phi,k} \big\rvert_{k\rightarrow0} 
    \simeq& \left( C_1 T + C_2\right)^{1/2},
    \end{split}
\end{equation}
where $C_1$ and $C_2$ are constants. 
This temperature dependence corresponds to the expected one in the cyan region of the phase diagram shown in Fig.~\ref{fig:CriticalFan}, where the exponent of the thermal length is $a=2$.

\section{Quasiparticle weight for chiral Ising and chiral XY models}\label{app:QpIXY}

\begin{figure}[h]
\centering
\includegraphics[width=0.85\linewidth]{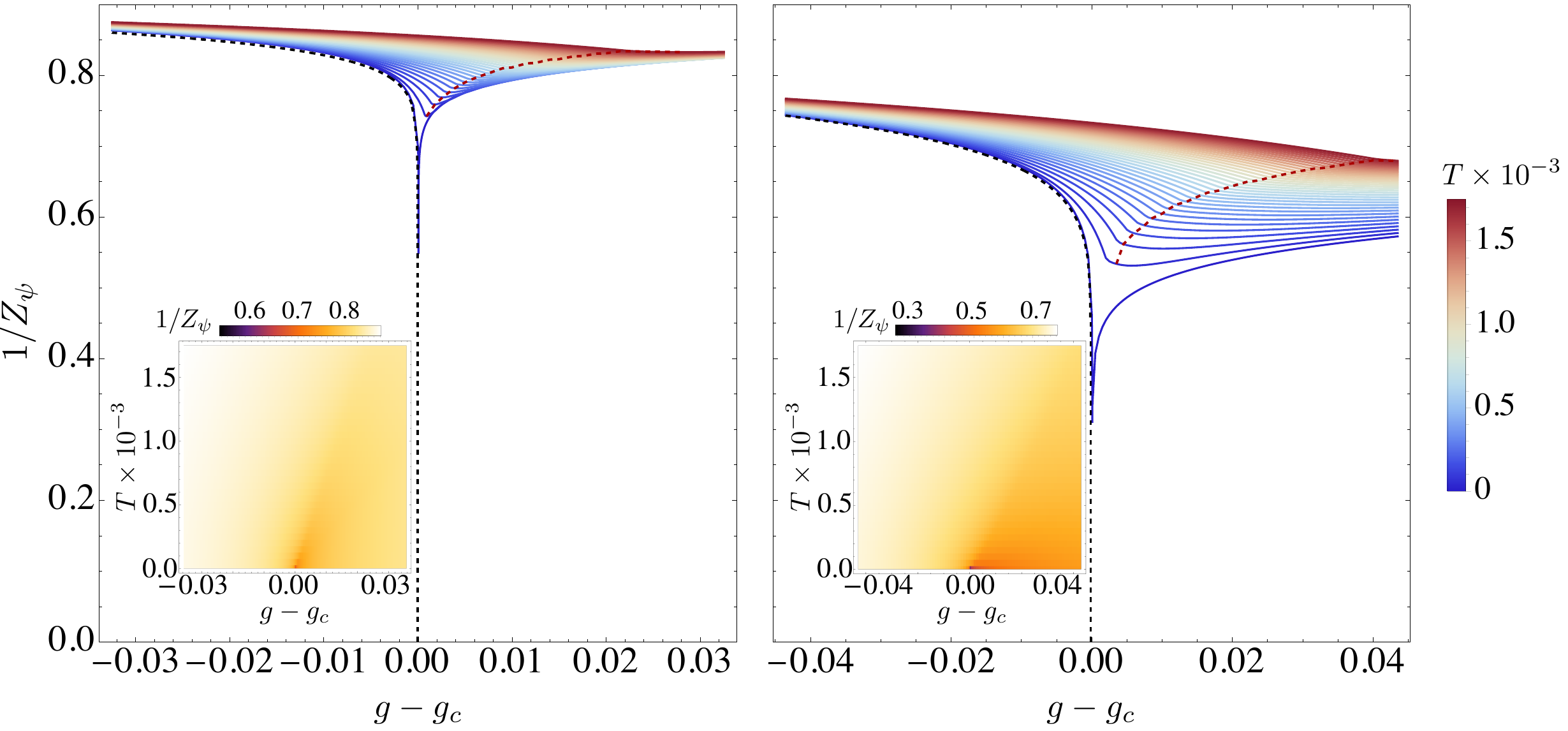}
\caption{\textbf{Fermionic quasiparticle weight} given by the inverse of the fermionic wave function renormalization $1/Z_{\psi,k}$ as a function of the ultraviolet coupling $g$ and temperature $T$ for the chiral Ising (left panel) and chiral XY (right panel) models with $N_\mathrm{f}=2$. The black dashed line corresponds to the fit provided in Eq.~\eqref{eq:HerbutFit}.
The red dashed line corresponds to the precondensation line.
Inset: fermionic quasiparticle weight as function of temperature and distance to the critical point.}
\label{fig:InvZpsi}
\end{figure}

In Fig.~\ref{fig:InvZpsi}, we present the fermionic quasiparticle weight for the chiral Ising and chiral XY models, which exhibit behavior
analogous to that observed for the chiral Heisenberg model in Fig.~\ref{fig:NDLH}. In both cases, we also observe a pronounced suppression
of the fermionic quasiparticle weight in the vicinity of the QCP, indicating non-Dirac liquid behavior.

\end{document}